\PassOptionsToPackage{utf8}{inputenc}
\documentclass{bioinfo}

\usepackage[table,xcdraw]{xcolor}
\usepackage[noend]{algpseudocode}
\usepackage{floatrow}
\usepackage{graphicx}
\usepackage{svg}
\usepackage{natbib}
\usepackage{algorithmicx,algorithm}
\usepackage{multirow}
\usepackage{amsmath}
\usepackage{url}
\usepackage{parskip}
\usepackage{xcolor}
\copyrightyear{2022} \pubyear{2022}

\def\method{{\fontfamily{lmtt}\selectfont RAGCell}}

\access{Advance Access Publication Date: Day Month Year}
\appnotes{Manuscript Category}
\begin{document}
\firstpage{1}

\subtitle{Gene expression}

\title[short Title]{RAGCell: Retrieval-Augmented Generation as Supervision for Versatile Single-cell Analysis}
\author[Liu \textit{et~al}.]{Tianyu Liu$^{1,8,\ast,\#}$, Fan Zhang$^{2,\#}$,  Jiayuan Chen$^{3,\#}$, Kun Wang$^{4}$, Haoxuan Li$^{5}$, Shengju Qian$^{2}$, Zhihong Zhu$^{5}$, Donghao zhou$^{2}$, Hao Wu$^{6}$, Ziheng Zhang$^{6}$, Zhenxi Lin$^{6}$, Xian Wu$^{6}$, Yefeng Zheng$^{7,\ast}$}

\address{$^{\text{\sf 1}}$Tsinghua University, Beijing, China,
$^{\text{\sf 2}}$The Chinese University of Hong Kong, Hong Kong SAR, China, 
$^{\text{\sf 3}}$The Ohio State University, Columbus, OH, USA,
$^{\text{\sf 4}}$Nanyang Technological University, Singapore,
$^{\text{\sf 5}}$Peking University, Beijing, China, 
$^{\text{\sf 6}}$Tencent Jarvis Lab, Shenzhen, China, 
$^{\text{\sf 7}}$Westlake University, Hangzhou, China,
$^{\text{\sf 8}}$Yale University, New Haven, CT, USA
}
\corresp{$^\ast$: To whom correspondence should be addressed.\\
\#: These authors contribute equally to this work.}

\history{Received on XXXXX; revised on XXXXX; accepted on XXXXX}

\editor{Associate Editor: XXXXXXX}

\abstract{\textbf{Motivation:} 
Single-cell foundation models (scFMs) are transforming computational biology by enabling generalizable, task-agnostic representations for versatile single-cell analysis. Despite their progress in facilitating rapid deployment for downstream tasks, off-the-shelf scFMs still have some overlooked concerns: (I) (\textit{Pretraining Cost.}) Pretrain-based scFMs necessitate pretraining on a vast volume of cells, rendering it draining resources in applications. (II) (\textit{Heterogeneous Gap.}) Large Language Models (LLM)-based scFMs ignore the tremendous heterogeneous gap between LLM textual and raw cellular spaces, leading to insufficient capability when facing downstream tasks.\\
\textbf{Results:} 
To this end, we introduce \method{}, a versatile single-cell analysis framework that achieves a double-win in both \textbf{cost-effectiveness} and \textbf{high performance}. The success of \method{} lies in two key aspects: Leveraging LLMs to construct cell-level and feature-level knowledge databases, which serve as supervision signals for training the cell model and significantly reduce the training cost ($>$pretrain-based scFMs). Aligning cell representations with text embeddings from the bi-level knowledge databases, enabling knowledge transfer from textual spaces to cellular spaces and effectively mitigating the heterogeneous gap ($>$LLM-based scFMs). Through extensive experiments on six downstream single-cell analysis tasks, we demonstrate that \method{} achieves outstanding performance compared to state-of-the-art scFMs while operating at less than $\sim$1/10 the cost of pretrain-based scFMs. \\
\textbf{Availability:} The source code is available at our supplementary file. Embeddings can be found in \url{https://sites.google.com/yale.edu/scelmolib/home}.\\
\textbf{Contact:} \href{tianyu.liu@yale.edu}{tianyu.liu@yale.edu}\\
\textbf{Supplementary information:} Supplementary data are available at \textit{Bioinformatics}
online.}
\maketitle

\section{Introduction}

Recent advances in machine learning and large language models (LLMs)~\cite{devlin2018bert,brown2020language,ouyang2022training,achiam2023gpt,touvron2023llama} have greatly facilitated single-cell analysis. Several single-cell foundation models (scFMs) have been proposed to obtain task-agnostic cell representations that generalize well to specific downstream single-cell analysis tasks.
These scFMs can be broadly divided into two groups: pretrain-based scFMs and LLM-based scFMs.
Pretrain-based scFMs typically leverage a vast volume of single-cell RNA sequence (scRNA-seq) data for large-scale pretraining, while LLM-based scFMs usually construct cell representations from LLM embeddings.
Although these scFMs have made significant progress for versatile single-cell analysis, there are several limitations that cannot be neglected. 
\textit{Firstly}, the success of pretrain-based scFMs depends on the pretraining process over a large volume of cells, which is time-consuming and resource-intensive in practice. Most existing scFMs still require large-scale pretraining on massive single-cell corpora before they can be used in a zero-shot manner. Thus, while users may not need to fine-tune or retrain the model for every new task, the initial development of such models remains computationally expensive and data intensive. To obtain scalable embeddings for single-cell data from different resources, pretrain-based scFMs often define gene vocabularies for data tokenization, which requires genomics knowledge from human experts.
Despite their outstanding performance, the costs associated with them are challenging to bear.
Obtaining expert priors is difficult and demands labor-intensive efforts, consequently leading to a decrease in training efficiency.
\textit{Secondly}, LLM-based scFMs often ignore heterogeneity between raw cellular and LLM textual spaces.
The LLM textual space is constructed based on pretraining in natural language, whereas the cellular space is dedicated to modeling biological data, leading to inherent significant gaps and heterogeneity.
Although LLM-based scFMs are cost-effective, their performance often falls short, especially in finetuning scenarios.


Inspired by the recent progress of LLM agents ~\cite{wang2024survey,durante2024agent} and the associated techniques like retrieval-augmented generation (RAG)~\cite{lewis2020retrieval,zhao2024retrieval} and its applications in biological tasks ~\cite{yu2025scrag}, we present \method{}, a versatile single-cell analysis framework that achieves a double-win in both \textbf{cost-effectiveness} and \textbf{high performance}. Here we define RAG as a computational paradigm. RAG is the decoupling of parametric knowledge (stored in model weights) from non-parametric knowledge (stored in an external, retrievable index), with the latter dynamically queried to augment downstream generation or prediction. \method{} follows precisely this paradigm: rather than encoding all biological knowledge in model parameters, we maintain an external index of representations that are learned at training time to condition the model's output in the inference time. The retrieval mechanism, not the textual nature of the output, is the conceptual core of RAG. The retrieved representations are not only used for alignment in the contrastive-learning sense, but also injected into the model to augment its prediction for a given query cell. This is functionally parallel to a standard RAG pipeline where retrieved documents are concatenated into an LLM's context to inform its output. Our motivation is to reduce this upstream pretraining burden by leveraging the knowledge already encoded in general-purpose language models and converting cell- or feature-level biological information into meaningful textual representations.
Different from existing LLM-based scFMs that directly leverage LLM embeddings for single-cell data modeling, \method{} firstly generates cell-level and feature-level knowledge databases through LLMs. We check the responses to ensure that they describe general cell-type and feature biology rather than the downstream prediction itself.
For cell-level text descriptions, the omics and cell type information are included in a prompt template as queries, and corresponding responses from LLMs are stored in the database.
For feature-level text descriptions, the specific functions of genes~\cite{chen2019single}, peaks~\cite{zeng2024deciphering}, and proteins~\cite{wu2012single} are preserved in another database.
Then, we employ a patch-based light-weighted Transformer \cite{vaswani2017attention} as the cell model to obtain cell representations.
Afterward, the text descriptions from these databases are retrieved and projected into a new space to perform cell-text alignment~\cite{chen2020simple,chen2020big,chen2020improved,chen2021empirical,radford2021learning}, which aims to provide supervision for the cell model and align cell representations from the cellular space with knowledge from the well-defined LLM textual space.
After alignment, the cell model could be employed for various downstream single-cell analysis tasks under both finetuning and zero-shot settings.
Compared with existing scFMs, \method{} can model single-cell multi-omics data into a general framework and achieve superior performance on downstream tasks free of the need for extensive pretraining on extra single-cell data or any human expert priors.
In comparison to pretrain-based scFMs such as scBERT~\cite{yang2022scbert} and scGPT~\cite{cui2024scgpt}, \method{} can achieve more than a \underline{\textbf{tenfold}} reduction in pretraining costs.
In addition, since omics information is included in the cell-level database, \method{} is scalable to both single-omics and multi-omics data analysis tasks, such as cell type annotation, batch effect correction, multi-omics data integration, rare cell type annotation, and drug sensitivity prediction.
The superiority of \method{} against a range of cutting-edge scFMs is fully demonstrated through comprehensive experiments on many downstream single-cell analysis tasks and datasets.
To summarize, the main contributions of this paper are threefold:
\begin{enumerate}
    \item \textbf{\textit{Novel Perspective.}} We identify the key limitations of pretrain-based and LLM-based scFMs, and then provide a novel perspective to reduce the training cost of single-cell analysis framework while maintaining its capability by incorporating LLM priors as supervision signals.
    \item \textbf{\textit{Double-Win Framework.}} We present \method{}, a versatile framework that is based on RAG and empowered by LLMs. Leveraging LLM-generated information as supervision, \method{} achieves a double-win in high performance and cost-effectiveness for single-cell analysis.
    \item \textbf{\textit{Multifaceted Validation.}} We conduct comprehensive experiments on five downstream single-cell analysis tasks and demonstrate the superiority of \method{} against many state-of-the-art scFMs.
\end{enumerate}

\section{Related Work}

\subsection{Single-cell Analysis}
Single-cell analysis seeks to characterize cellular states across various omics modalities, such as RNA~\cite{saliba2014single,kolodziejczyk2015technology}, DNA~\cite{karemaker2018single,evrony2021applications}, and proteins~\cite{wu2012single,suman2015single}. As a foundational technique in computational biology, it has wide-ranging applications in healthcare~\cite{hong2019single,rajewsky2020lifetime} and medicine~\cite{lim2020advancing,paik2020single}. For example, single-cell analysis enables the identification of cellular heterogeneity, offering insights into the complexity of tissues and organs. It also plays a key role in elucidating disease progression, thereby supporting the discovery of novel therapeutic targets and advancing our understanding of disease mechanisms.
Numerous computational tasks have been developed for analyzing single-omics or multi-omics single-cell data, such as cell type annotation~\cite{jiang2023scatanno,hou2024assessing}, batch effect correction~\cite{tran2020benchmark,fei2020scbatch}, and multi-omics data integration~\cite{lance2022multimodal,cao2022multi}.

\begin{figure*}[t]
    \centering
    \includegraphics[width=0.8\linewidth]{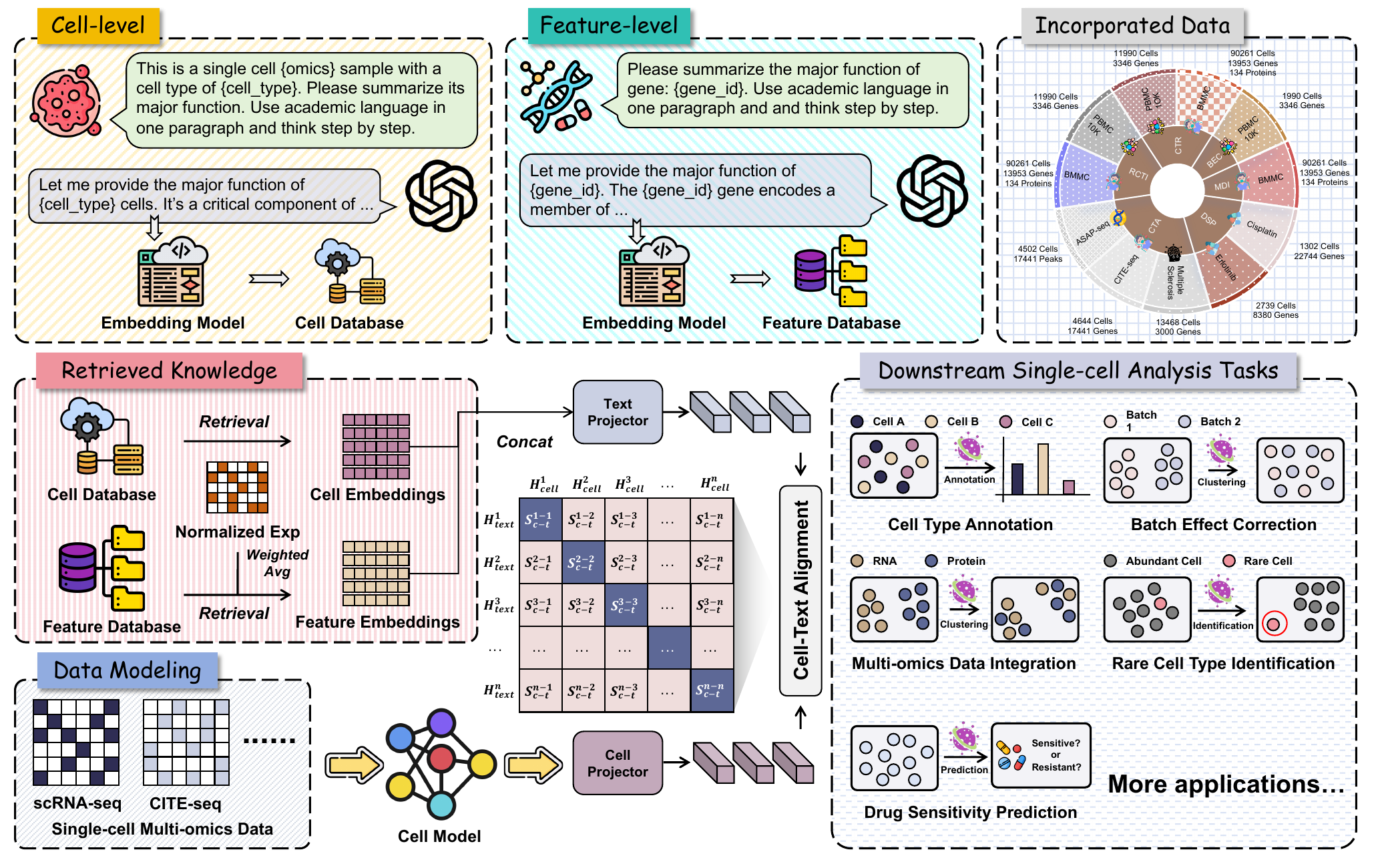}
    \caption{\textbf{An overview of \method{}.} \method{} utilizes multi-source single-cell data to construct both cell-level and feature-level knowledge databases. Subsequently, it retrieves bi-level knowledge as supervision signals for training the cell model. Following alignment of cell representations with the LLM textual space, the pretrained cell model can be applied to various single-cell analysis tasks.}
    \label{fig: framework}
\end{figure*}

\subsection{Foundation Models for Single-cell Biology}

With the rapid advancement of generative artificial intelligence~\cite{devlin2018bert,brown2020language,zhao2024langcell} and LLMs~\cite{touvron2023llama,achiam2023gpt}, numerous FMs have been proposed for single-cell biology.
For instance, scBERT~\cite{yang2022scbert} applies a bag-of-words strategy to discretize gene expression based on transcription frequencies and incorporates predefined Gene2Vec embeddings to tokenize single cells. It utilizes a Performer encoder~\cite{choromanski2020rethinking} to learn cell representations and conducts self-supervised pretraining on over one million scRNA-seq samples.
scGPT~\cite{cui2024scgpt} tokenizes data by incorporating gene expression values, gene tokens, and condition tokens. It employs a Transformer model trained autoregressively with an attention masking mechanism. 
In parallel, models such as GenePT~\cite{chen2023genept} and scELMo~\cite{liu2023scelmo} leverage LLMs to generate gene-level embeddings, which are then integrated with raw single-cell sequences to enhance representation learning and task performance. LangCell \cite{zhao2024langcell} aims to enhance single-cell models by combining text descriptions and pre-training. In comparison with these methods, our \method{} utilizes RAG techniques to achieve superior performance in versatile single-cell analysis with limited budgets.

\section{The \method{} Framework}

The proposed framework \method{} comprises a cell model and two associated knowledge databases. For each sample, we first create text descriptions with rich external information at both the cell and feature levels. These descriptions are then vectorized to obtain embeddings via embedding models and stored in databases for future use. Subsequently, we retrieve the generated text embeddings to perform cell-text alignment. In this stage, knowledge from LLM-based databases is utilized to guide the training process of a cell model, with the objective of injecting priors from the LLM textual space into the cellular space. 
Once the alignment is complete, the cell model can be employed to execute various downstream single-cell analysis tasks.
An overview of our framework is provided in Figure \ref{fig: framework}.

Regarding novelty, \method{}'s main contribution lies in its dual-granularity contrastive learning strategy, which embeds biological semantics directly into the gene/protein/peak representation space at both the cell and feature levels, presenting as a fundamentally different inductive bias from prior cell-language models. This is enabled by a purpose-built patch-based encoder whose patching scheme is designed to interact with feature-level supervision, as patches inherit aggregated descriptions from their constituent features. Therefore, the architecture and the supervision signal are mutually reinforcing rather than borrowed independently, which is also not trivial. The resulting model unlocks capabilities that the individual components cannot achieve in isolation, including zero-shot generalization to features unseen during training, cross-modality transfer, and biologically interpretable representations, as demonstrated in our applications and in the additional comparisons against other baselines.

\paragraph{Cell Model.} 
Since there is no need for extensive pretraining, we employ a lightweight Transformer model~\cite{dosovitskiy2020image} as the cell model. For a cell sample $\mathbf{X}\in \mathcal{R}^{1 \times N_f}$, we begin by splitting it into $M$ patches, and each patch is a $P$ dimensional vector where $N_f = M \times P$. Then we add a classification token $\mathbf{X}_\text{cls}$ to learn global representations and a 1D positional encoding $\mathbf{E}_\text{pos}$ to capture the relative position information:
\begin{equation}
    \mathbf{X}_t=[\mathbf{X}_\text{cls}; \mathbf{X}_p^1\mathbf{E}; \mathbf{X}_p^2\mathbf{E}; ... ; \mathbf{X}_p^M\mathbf{E}]+\mathbf{E}_\text{pos},
\end{equation}
where $\mathbf{X}_p^i$ and $\mathbf{E}$ represent patch embedding and linear transformation, respectively.
Afterward, we could employ a standard Transformer model~\cite{vaswani2017attention} to capture the cell representations:
\begin{equation}
    \mathbf{Z}_\text{cell} = \texttt{Transformer}(\mathbf{X}_t).
\end{equation}
Then, $\mathbf{Z}_\text{cell}$ can be utilized to perform cell-text alignment and downstream tasks.

\paragraph{LLMs.} 
LLMs are used to generate cell-level and feature-level text descriptions and to extract bi-level text embeddings, which introduce prior knowledge into the entire framework and thus eliminate the need for pretraining on large-scale data. In this paper, we opt for the GPT-4o mini model to generate cell-level text descriptions and the `text-embedding-3-large' model to extract embeddings. For feature-level text descriptions, we use the GPT-3.5 model to obtain the specific function for each gene or protein and the `text-embedding-ada-002' model to extract embeddings. To minimize hallucination and ensure biological accuracy in LLM-generated feature descriptions, four strategies are employed: context-aware generation that explicitly specifies modality and task setting to reduce entity ambiguity; post-generation validation against curated databases (with NCBI \cite{schoch2020ncbi}) to flag mismatched or inconsistent descriptions for regeneration or removal; manual inspection of high-impact examples used in figures or case studies; and treating generated descriptions as auxiliary rather than ground-truth supervision, with robustness checks and ablation analyses to assess sensitivity to individual descriptions.

\paragraph{Bi-level Knowledge Databases Construction.}
The core of our framework is to construct cell-level and feature-level knowledge databases. For each cell sample, we consider multiple aspects of information to be crucial for constructing high-quality representations, including omics information, cell type information, and specific feature information (such as gene or protein). Therefore, it is necessary to include all these types of information in the text descriptions.
First, we apply a text template: ``This is a single-cell $<$omics$>$ sequence sample with a cell type of $<$cell type$>$. Please summarize its major function. Use academic language in one paragraph and think step by step.". We then fill in the omics and cell type information for each cell into this template. Next, we utilize an embedding model (EM) to obtain text embeddings. The entire process can be formulated as:
\begin{equation}
    \mathbf{Z}_\text{text}^{c} = \texttt{EM}(\texttt{LLM}(\texttt{Prompt}(Q^{c}))),
\end{equation}
where $Q^{c}$ denotes cell-level prompts and $\mathbf{Z}_\text{text}^{c}\in \mathcal{R}^{1 \times F_1}$ represents the corresponding cell-level $F_1$ dimensional text embeddings.
Next, we follow previous works GenePT \cite{chen2023genept} and scELMo \cite{liu2023scelmo}, to obtain feature-level text embeddings. There are generally two ways to acquire feature-level text descriptions: using text descriptions from the NCBI database as prompts or using human-designed prompts in a dialogue with an LLM. Here, we illustrate the second approach as an example.
We begin by using a prompt following ~\cite{liu2023pre,jia2022visual,ekin2023prompt} to ask the LLM for specific feature information (using a gene as an example): ``Please summarize the major function of gene: $<$gene$>$. Use academic language in one paragraph and think step by step.". 
Subsequently, we utilize the LLM's response to this query as the feature-level text descriptions and obtain its corresponding embeddings. This process can be formulated as:
\begin{equation}
    \mathbf{Z}_\text{text}^{f_i} = \texttt{EM}(\texttt{LLM}(\texttt{Prompt}(Q^{f_i}))),
\end{equation}
where $Q^{f_i}$ denotes the query for each feature and $\mathbf{Z}_\text{text}^{f_i}\in \mathcal{R}^{1 \times F_2}$ represents the corresponding $F_2$ dimensional text embeddings. For a cell sample with $N_f$ features, we could stack $\mathbf{Z}_\text{text}^{f_i}\in \mathcal{R}^{1 \times F_2}$ for each feature and obtain the final feature-level text embeddings $\mathbf{Z}_\text{text}^{f}\in \mathcal{R}^{N_f \times F_2}$.
For each specific dataset, we can store these cell-level and feature-level embeddings in two databases. During the training process, the corresponding embeddings can be retrieved to provide supervision signals.

\paragraph{Cell-Text Alignment.}
After constructing bi-level knowledge databases, we then transfer knowledge from the databases to the cell model by retrieving relevant information. To begin, we need to concatenate the cell-level and feature-level text embeddings:
\begin{equation}
    \mathbf{Z}_\text{text}^{f'} = \mathbf{X} \times \mathbf{Z}_\text{text}^{f}, \quad
    \mathbf{Z}_\text{text} = \texttt{Concat}(\mathbf{Z}_\text{text}^{c},\mathbf{Z}_\text{text}^{f'}),
\end{equation}
where $\mathbf{Z}_\text{text}^{f'} \in \mathcal{R}^{1 \times F_2}$ and $\mathbf{Z}_\text{text} \in \mathcal{R}^{1 \times (F_1+F_2)}$.
Then, two separate multilayer perceptrons (MLPs) are utilized to project cell representations and text embeddings into new low-dimensional spaces:
\begin{equation}
    \mathbf{H}_\text{cell} = \texttt{MLP}(\mathbf{Z}_\text{cell}), \quad
    \mathbf{H}_\text{text} = \texttt{MLP}(\mathbf{Z}_\text{text}),
\end{equation}
where $\mathbf{H}_\text{cell} \in \mathcal{R}^{1 \times D}$ and $\mathbf{H}_\text{text} \in \mathcal{R}^{1 \times D}$ have same code length. Following prior works~\cite{chen2020simple,he2020momentum,radford2021learning,xiong2023scclip}, we set $D=128$ by default.
After getting $\mathbf{H}_\text{cell}$ and $\mathbf{H}_\text{text}$ for each cell, we then perform cell-text alignment via instance-level matching:
\begin{equation} 
L_\text{C2T} = -\frac{1}{B} \sum_{i}^B log \frac{exp(cos(\mathbf{H}_\text{cell}^i, \mathbf{H}_\text{text}^i))}{\sum_{j=1}^B exp(cos(\mathbf{H}_\text{cell}^i, \mathbf{H}_\text{text}^j))},
\end{equation}
\begin{equation} 
L_\text{T2C} = -\frac{1}{B} \sum_{i}^B log \frac{exp(cos(\mathbf{H}_\text{text}^i, \mathbf{H}_\text{cell}^i))}{\sum_{j=1}^B exp(cos(\mathbf{H}_\text{text}^i, \mathbf{H}_\text{cell}^j))},
\end{equation}
where $B$ denotes the number of samples within a mini-batch, $exp(\cdot)$ represents exponential transformation, and $cos(\cdot)$ represents cosine similarity.
The final loss objective can be formulated as:
\begin{equation}
    L = \frac{1}{2}(L_\text{C2T}+L_\text{T2C}).
\end{equation}
The loss function ensures that the distance between cell-text pairs is minimized in the representation space, while the distance between non-paired cell and text representations is maximized. Through pretraining on specific datasets, we inject semantic knowledge from LLMs into the cell model. 
After training, the cell model can be employed for many downstream single-cell analysis tasks.

\paragraph{Downstream Single-cell Analysis Tasks.}
We evaluate the performance of \method{} on several downstream single-cell data analysis tasks. For tasks like cell type annotation and drug sensitivity prediction, we first leverage a classification head to obtain the predictions for each cell:
\begin{equation}
    \mathbf{P}_\text{cell} = \texttt{MLP}(\mathbf{H}_\text{cell}).
\end{equation}
Then we finetune the cell model and classification head on specific datasets with cross-entropy loss:
\begin{equation}
    L_\text{cls} = \texttt{CE}(\mathbf{P}_\text{cell}, Y_\text{cell}),
\end{equation}
where $Y_\text{cell}$ denotes the corresponding cell type labels or drug sensitivity labels.
For batch effect correction and multi-omics data integration tasks, we directly employ the cell representations for evaluation, without the need for finetuning.
For the rare cell type identification task, we employ the SOTA method scCAD~\cite{xu2024sccad} as a baseline, which starts from clustering. Our modified scCAD is based on embeddings from \method{}, and is a purely zero-shot method. Then we replace the original cell embeddings with our cell representations. All other algorithm settings are kept consistent.

\paragraph{Inference-time retrieval and representation fusion.} At inference time, RAGCell uses the encoded cell representation ($H_{\mathrm{cell}}$) as a query to retrieve relevant textual knowledge from the external database. Specifically, all database entries are encoded in advance using the text encoder and stored in a searchable embedding index, and cosine similarity is used to rank their relevance to the query cell. The top-(k) text representations are selected and aggregated using similarity-normalized attention weights, yielding a retrieved knowledge representation ($H_{\mathrm{ret}}=\sum_{i=1}^{k}\alpha_i H_{\mathrm{text}}^{(i)}$), where $\alpha_i=\operatorname{softmax}(\operatorname{sim}(H_{\mathrm{cell}},H_{\mathrm{text}}^{(i)}))$. The retrieved representation is then fused with the original cell representation through a learnable projection, ($H_{\mathrm{aug}}=\operatorname{MLP}([H_{\mathrm{cell}};H_{\mathrm{ret}}])$), and ($H_{\mathrm{aug}}$), rather than $H_{\mathrm{cell}}$ alone, is provided to the downstream prediction head. The retrieval database and text embeddings are constructed without using test-set labels, and model parameters remain fixed during inference. This is implemented in the training process.

\section{Experiment}

\begin{figure*}[!t]
    \centering
    \includegraphics[width=0.8\linewidth]{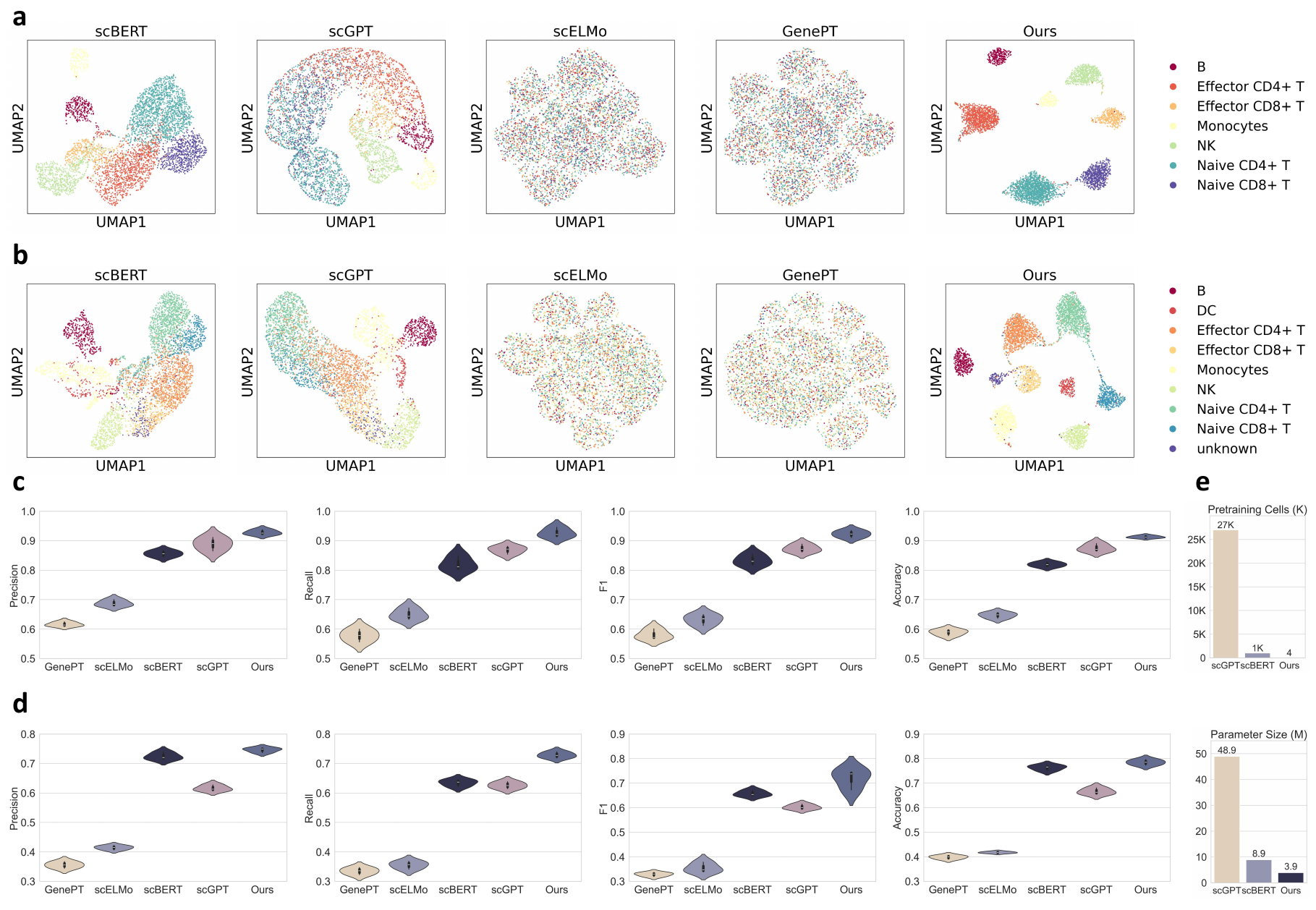}
    \caption{\textbf{Cell Type Annotation Results.} \textbf{a.} UMAP plot of the embeddings finetuned by five methods on CITE-seq data. \textbf{b.} UMAP plot of the embeddings finetuned by five methods on ASAP-seq data. \textbf{c.} Precision, Recall, F1, and Accuracy comparisons of different methods on CITE-seq data. \textbf{d.} Precision, Recall, F1, and Accuracy comparisons of different methods on ASAP-seq data. \textbf{e.} Cost comparisons.}
    \label{fig: cta}
\end{figure*}

\subsection{Cell Type Annotation}
Cell type annotation~\cite{cao2020scsa,chen2022cell,chen2023Transformer,shao2021scdeepsort} is known as a crucial task in single-cell analysis. 
However, most existing scFMs focused exclusively on scRNA-seq data and lacked validation across other omics modalities. 
To address this gap, we evaluated \method{} against other scFMs using CITE-seq~\cite{stoeckius2017simultaneous}, ASAP-seq~\cite{mimitou2021scalable} and Liver data~\cite{lin2020scclassify,aizarani2019human}. 
For fair comparisons, we randomly partitioned CITE-seq and ASAP-seq into three subsets: 80\% for training, 10\% for validation, and 10\% for testing. The Liver dataset is tested by using different batch for training/testing. The liver evaluation uses two independent studies, \cite{macparland2018single} (GSE115469) and \cite{aizarani2019human} (GSE124395), and is therefore a cross-study rather than a cross-batch transfer.  Their annotations were harmonised to seven cell types shared by both (B/plasma, cholangiocytes, hepatocytes, LSEC, macrophages, macrovascular/portal, T/NK), and both datasets were restricted to the 17,146 genes common to them. Each study was split 80/10/10 into training, validation and test partitions, stratified by cell type; a classifier is trained on one study and evaluated on the held-out test partition of the other, in both directions. Fig. \ref{fig: cta} (a) and (b) illustrate cell embeddings finetuned by the five methods. The UMAP visualizations~\cite{mcinnes2018umap} demonstrated that \method{} achieved superior separation of distinct cell types for both CITE-seq and ASAP-seq data, confirming its effectiveness for multi-omics cell type annotation.
We subsequently conducted comprehensive quantitative comparisons using four evaluation metrics.
As depicted in Fig.~\ref{fig: cta} (c) and (d), pretrain-based scFMs (scBERT \cite{yang2022scbert}, scGPT \cite{cui2024scgpt}) outperform LLM-based scFMs (GenePT \cite{chen2023genept}, scELMo \cite{liu2023scelmo}) by a large margin, consistent with expectations given the computationally intensive pretraining required for Transformer-based models. Remarkably, \method{} consistently surpassed all other scFMs across every metric. We also compare more baseline based on both Liver and ASAP-seq datasets to test \method{} with more biological context. Figure \ref{fig: morebench_asap} shows that \method{} performs better than Cell2Sentence, Geneformer, and UCE, and comparable with scFoundation. Training scFMs requires larger resources, and thus \method{} works more efficiently. Figures \ref{fig: morebench_asap} (a) and (b) show the cross-dataset performances based on the Liver dataset, where \method{} still presents as top performers. Therefore, \method{} still servers as a strong method for cell-type annotation under more complicated scenarios. Based on the ASAP-seq dataset, we also test whether using different LLMs to generate feature embeddings will affect the results obviously or not. We compare the embeddings generated based on descriptions from GPT 3.5 and GPT 4o-mini, as shown in Figure \ref{fig: morebench_gpt}, there is no obvious difference between these two methods.

To test whether the gains hold beyond a single tissue and at a larger scale, we added two datasets. Tabula Sapiens Immune~\cite{the2022tabula} is the immune compartment of the Tabula Sapiens atlas, drawn across multiple organs and donors, and Fetal Bone Marrow~\cite{jardine2021blood} is a developmental haematopoietic atlas; the former therefore probes generality across tissues and the latter across developmental stages. From each atlas we retained the cell types represented by at least 100 cells and then drew a uniform random subsample of 100,000 cells without replacement under a fixed seed, giving 46 types over 58,604 genes and 40 types over 32,922 genes respectively. We use the similar split setting, and results are shown in Figures \ref{fig: largescale} (a) and (b). Both label sets are considerably finer than those used elsewhere in this study and both are strongly imbalanced, which is what makes them informative: in Fetal Bone Marrow the largest type accounts for 18,480 cells while the rarest sit near the 100-cell floor. On Tabula Sapiens Immune RAGCell attains the best score on all four metrics, with a macro-F1 of 0.723 against 0.691 for scFoundation and 0.663 for UCE, and the margin is consistent across the supervised, neighbourhood and clustering metrics rather than confined to one of them. On Fetal Bone Marrow the picture is mixed: RAGCell is second on accuracy but third on macro-F1, behind scFoundation and UCE. The discrepancy is informative rather than incidental. Accuracy and the neighbourhood metric are dominated by the abundant erythroid and precursor populations, on which RAGCell is competitive, whereas macro-F1 weights every type equally and is therefore governed by the rarest types, which here are lymphoid subsets such as natural killer and CD8 T cells sitting near the 100-cell floor.

It is worth noting that \method{} achieved this performance without requiring supplemental pretraining data, supported by the comparison for GPU resource and single-cell data corpus size (Fig.~\ref{fig: cta} (e)). Its bi-level knowledge databases were constructed exclusively from training sets of the CITE-seq and ASAP-seq datasets, contrasting sharply with pretraining-based scFMs (e.g., scBERT, scGPT) that relied on massive pretraining corpora (millions to tens of millions of cells). These results indicate that \method{} effectively synthesized strengths from both paradigms: leveraging external knowledge from LLM-based databases enabled state-of-the-art performance with manageable training costs.

\begin{figure*}[ht]
    \centering
    \includegraphics[width=0.8\linewidth]{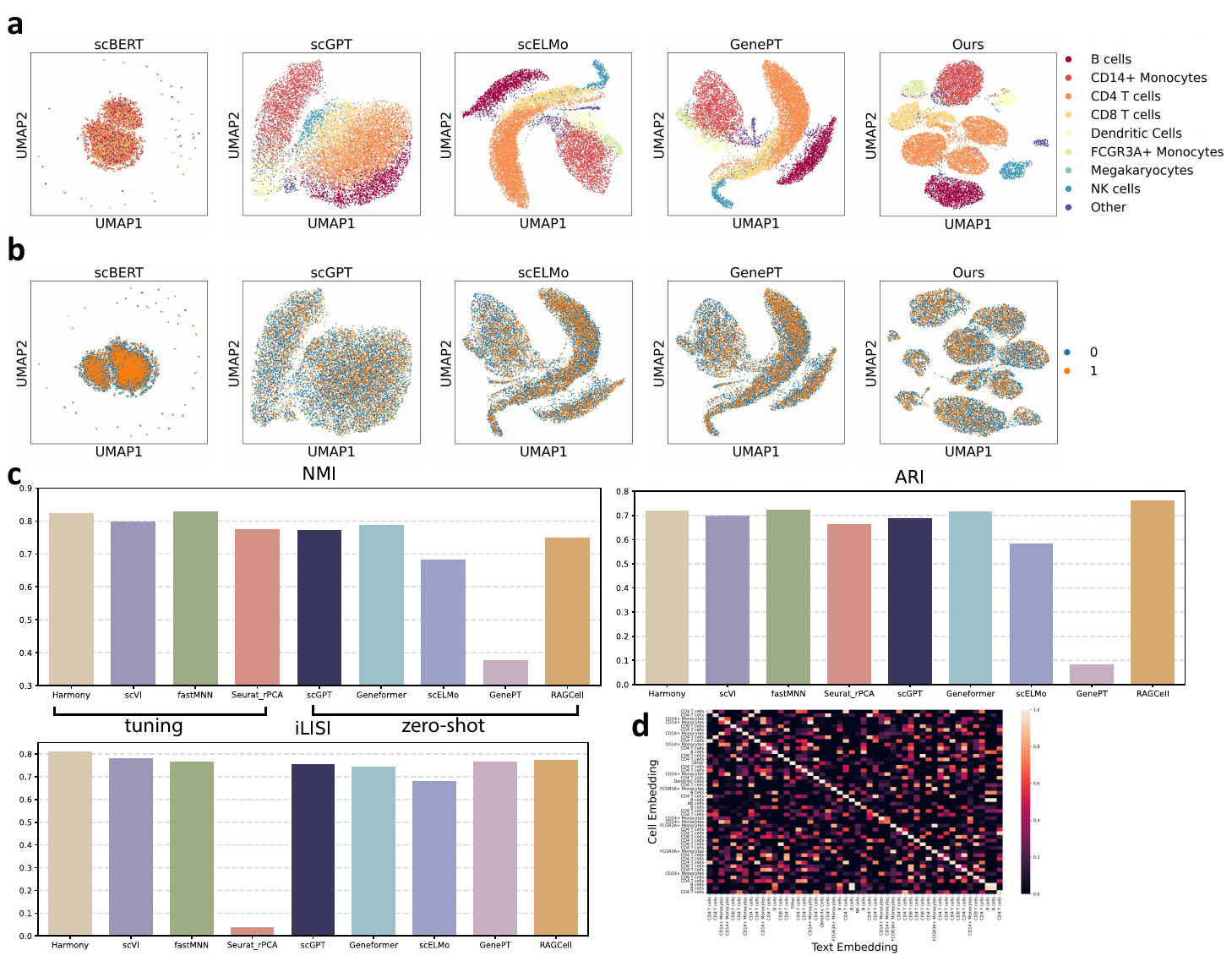}
    \caption{\textbf{Batch Effect Correction Results on PBMC 10K under Zero-shot Settings.} \textbf{a.} UMAP plot of the embeddings by five methods across cell types. \textbf{b.} UMAP plot of the embeddings by five methods across batches. \textbf{c.} Batch Effect Correction performance comparisons among five methods. We list both tuning methods and zero-shot methods. \textbf{d.} Heatmap visualizations of similarity between cell and text embeddings on randomly selected 50 cells.}
    \label{fig: batch_effect_pbmc}
\end{figure*}

\begin{figure*}[ht]
    \centering
    \includegraphics[width=0.8\linewidth]{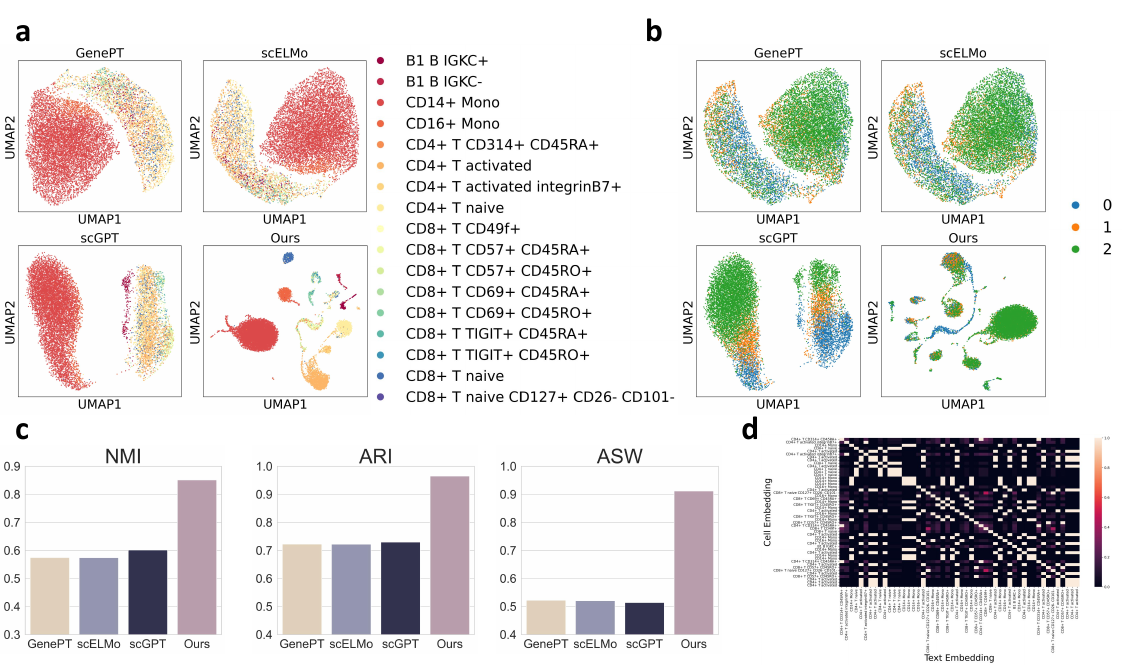}
    \caption{\textbf{Multi-omics Data Integration Results on BMMC data under Zero-shot Settings.} \textbf{a.} UMAP plot of the embeddings by different methods across cell types. \textbf{b.} UMAP plot of the embeddings by different methods across batches. \textbf{c.} Cluster performance comparisons. \textbf{d.} Heatmap visualizations of similarity between cell and text embeddings on randomly selected 50 cells.}
    \label{fig: multiomics_bmmc}
    \vspace{-2mm}
\end{figure*}

\subsection{Batch Effect Correction}

Batch effect correction represents another fundamental challenge in single-cell analysis. To assess whether \method{} effectively addresses batch effect issues~\cite{tran2020benchmark,goh2017batch,fei2020scbatch}, we conducted experiments on the PBMC 10K dataset~\cite{gayoso2022python} containing samples from two distinct batches and consider metrics which can evaluate both cell-type preservation and batch effect correction. All metrics were computed with the standard scIB toolkit (v1.1.6) \cite{luecken2022benchmarking} under identical settings across all methods. Generalization capability was evaluated under zero-shot settings, with \method{} pretrained on around 65,000 cells from the NeurIPS 2021 Multimodal Single-Cell Data Integration competition~\cite{luecken2021sandbox}. This approach demonstrates significantly higher GPU resource efficiency compared to scBERT and scGPT.
The visualization results in Fig.~\ref{fig: batch_effect_pbmc} (a) reveal that \method{} generates more discriminative cell-type clusters than alternative scFMs. 
scBERT exhibited the weakest performance among these scFMs, failing to form distinct clusters. This limitation stems from its masked value reconstruction pretraining strategy, which is no help for dealing with batch effects or forming discriminative clusters.
While GenePT, scELMo, and scGPT produced partial clusters, their cluster counts fell substantially below the true number of cell types.
In contrast, \method{} generated clusters corresponding closely to the actual cell type count with well-separated cluster centers, indicating superior batch integration capability. This enhancement stems from integrating prior knowledge retrieved from LLM-based databases, effectively suppressing batch-specific variations.
The results in Fig.~\ref{fig: batch_effect_pbmc} (b) further confirm substantial overlap between batches within \method{}'s latent space. Collectively, these results demonstrate \method{}'s effectiveness in batch effect correction and clustering improvement for PBMC 10K data.
Quantitative validation results in Fig.~\ref{fig: batch_effect_pbmc} (c) using standardized metrics (NMI, ARI for cell-type preservation, and iLISI score for batch effect removal~\cite{luecken2022benchmarking}) indicate that \method{} consistently outperforms baseline methods in batch effect correction. Despite a marginal ARI deficit, its dominant advantages in other metrics substantiate its robust capabilities. Moreover, \method{} serves as a zero-shot-based methods, and presents as a top performer under this category. We also illustrate ASW score comparisons (Figure \ref{fig: morebench_bec}), and the results are consistent.
In Fig.~\ref{fig: batch_effect_pbmc} (d), we can observe that cell-text alignment induces high similarity between embeddings of biologically similar cells, effectively disentangling biological signals from technical batch effects.

\begin{figure*}[ht]
    \centering
    \includegraphics[width=0.8\linewidth]{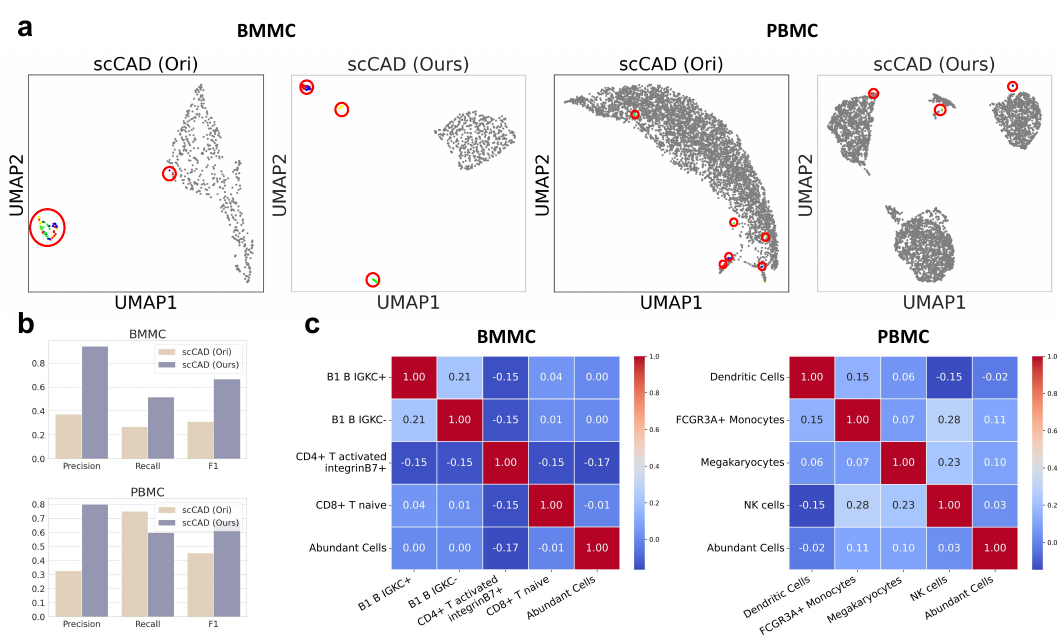}
    \caption{\textbf{Rare Cell Type Identification Results under Zero-shot Settings.} \textbf{a.} UMAP plot of the cell embeddings related to abundant and rare cell types. \textbf{b.} Quantitative performance comparison between the original cell embeddings and ours using the scCAD algorithm. \textbf{c.} Heatmap visualizations of similarity between average embeddings of rare cells and abundant cells.}
    \label{fig: rare_cell}
\end{figure*}

\subsection{Multi-omics Data Integration}
By leveraging an LLM to derive text embeddings capturing omic-specific information, the cell model within \method{} acquires inherent multi-omic representational capabilities during the alignment process. Consequently, \method{} excels at multi-omics data integration. To validate this capability, we utilized RNA and Protein modalities from the BMMC dataset~\cite{luecken2021sandbox} to evaluate zero-shot clustering performance, employing the same pretraining data as described in the previous section's batch effect correction experiment. Here LLM-produced embeddings are used to model both RNA and Protein modalities, with text descriptions correspondingly.
Fig.~\ref{fig: multiomics_bmmc} (a) visualizes cell embeddings from GenePT, scELMo, scGPT, and \method{}. While all methods distinguished CD14+ monocytes, other scFMs proved unable to resolve CD8+ T cells. LLM-based scFMs (GenePT, scELMo) exhibited highly overlapping embeddings for remaining cell types, suggesting insufficient discriminative power despite underlying gene expression differences. scGPT showed marginal improvement by disambiguating CD4+ T cells, yet remained incapable of distinguishing other cell populations. In contrast, \method{} successfully transferred semantic knowledge from the LLM to the cell model, yielding well-separated clusters for nearly all cell types and superior visualization quality.
Fig.~\ref{fig: multiomics_bmmc} (b) demonstrates \method{}'s robust batch integration, evidenced by the highest degree of inter-batch embedding overlap. Conversely, GenePT, scELMo, and scGPT persistently exhibited batch-specific structures, indicating inconsistent capture of biologically relevant features amidst technical noise.
Quantitative validation in Fig.~\ref{fig: multiomics_bmmc} (c) using NMI, ARI, and ASW metrics revealed \method{}'s dominant performance, with all scores exceeding 0.8. The remaining methods consistently fell below this threshold (scores < 0.8), further confirming \method{}'s superiority in multi-omics integration.
Finally, Fig.~\ref{fig: multiomics_bmmc} (d) illustrates elevated embedding similarity between biologically similar cells post-alignment. This correlation demonstrates that text embeddings provide effective semantic regularization, enabling \method{} to seamlessly integrate both RNA and protein data while simultaneously mitigating batch effects.

\subsection{Rare Cell Type Identification}

Advances in sequencing technologies have yielded vast quantities of scRNA-seq data~\cite{hwang2018single,cao2020searching}, which frequently contain both abundant and rare cell populations. Although rare cell types~\cite{travaglini2020molecular,wu2019advantages,kiselev2019challenges} exhibit low abundance, they play pivotal roles in biological processes such as disease pathogenesis and drug discovery. Consequently, rare cell identification has emerged as a critical challenge in single-cell analysis~\cite{jiang2016giniclust,jindal2018discovery,dong2020giniclust3}.
By leveraging prior knowledge from LLMs to derive high-quality cell representations, \method{} provides an effective framework for rare cell identification. We adopted the state-of-the-art scCAD algorithm~\cite{xu2024sccad} as our baseline, which initially clusters cells via principal component analysis (PCA). To evaluate \method{}'s capability, we replaced scCAD's original embeddings with \method{}'s zero-shot representations.
Fig.~\ref{fig: rare_cell} (a) (left) illustrates BMMC data visualizations. Whereas baseline results show abundant cells clustered distantly from most rare cells (with partial overlap), \method{} (right) achieves clear separation between abundant and rare populations while enhancing discrimination among rare cell subtypes. Similarly, for PBMC data (Fig.~\ref{fig: rare_cell} (a) (right)), baseline embeddings exhibit substantial mixing of rare and abundant cells, whereas \method{} maintains distinct inter-group distances and intra-group dispersion—significantly improving scCAD's rare cell identification capacity.
In Fig.~\ref{fig: rare_cell} (b), we selected three metrics to quantitatively compare the effectiveness of \method{} in enhancing the identification of rare cell types. The results show that, compared to the original cell embeddings, the embeddings obtained from our framework achieve improvements across all metrics on the BMMC dataset. On the PBMC dataset, while our recall score is slightly lower than that of the original embeddings, our performance on the other two metrics is superior.
Quantitative analysis in Fig.~\ref{fig: rare_cell} (b) using three metrics demonstrates \method{}'s consistent improvement across all measures on BMMC data. For PBMC data, while recall shows marginal reduction, \method{} outperforms baselines in two other metrics. Further validation in Fig.~\ref{fig: rare_cell} (c) reveals substantially reduced cosine similarity between abundant/rare cells and among distinct rare cell types, confirming \method{}'s discriminative power. We also include more baselines in the Figure \ref{fig: morebench_rarecelltypedetect}, which showcases that embeddings from other methods cannot surpass the settings introduced by \method{} especially based on F1 score, and thus \method{} can provide strong embeddings to identify rare cell types.

\begin{figure*}[!t]
    \centering
    \includegraphics[width=0.8\linewidth]{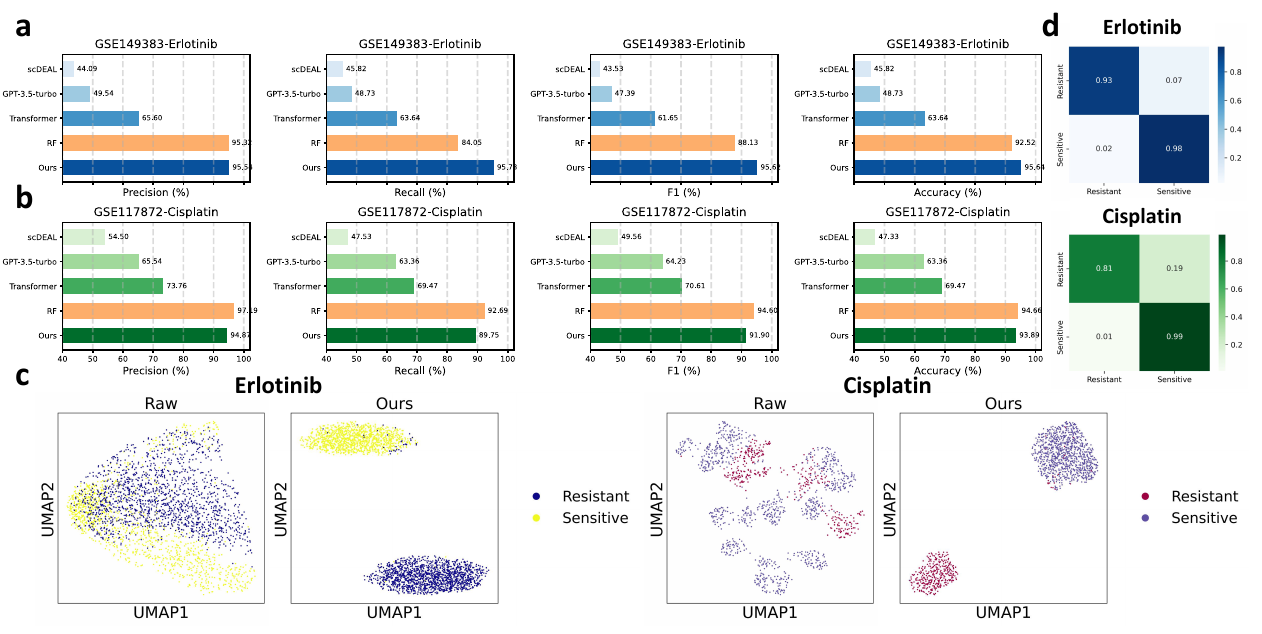}
    \caption{\textbf{Drug Sensitivity Prediction Results on Erlotinib and Cisplatin data.} \textbf{a.} Quantitative performance comparisons on Erlotinib data. \textbf{b.} Quantitative performance comparisons on Cisplatin data. \textbf{c.} UMAP plot of the raw embeddings and ours. \textbf{d.} Confusion matrix results.}
    \label{fig: dsp}
\end{figure*}

\subsection{Drug Sensitivity Prediction}

Accurate prediction of cellular drug sensitivity represents a critical challenge in precision biomedicine \cite{cortes2016current,ahmed2020network}. To evaluate \method{}'s capabilities, we conducted experiments on the GSE149383-Erlotinib \cite{aissa2021single} and GSE117872-Cisplatin \cite{sharma2018longitudinal,ravasio2020single,suphavilai2021predicting} datasets. The train/validation/test split is performed before any cell–text alignment or fine-tuning, and that both stages use only the training split, with no test-set sensitivity labels used for text database construction, alignment, model selection, or downstream training. Fig.~\ref{fig: dsp} (a) and (b) demonstrates \method{}'s superior performance over both specialist models (Transformer \cite{vaswani2017attention}, scDEAL \cite{chen2022deep}), generalist foundation models (GPT-3.5-turbo), as well as machine learning models such as Random Forest (RF). Across all four metrics, comparator methods scored below 0.8, while \method{} consistently exceeded 0.9, demonstrating significant improvements. This enhancement stems from our framework's ability to leverage inherent biological patterns within knowledge databases, strengthening cell-drug sensitivity associations. Due to the limitation of dataset number, we can only perform randomly and in-distribution split to create training/validation/testing datasets.
The visualization results in Fig.~\ref{fig: dsp} (c) further substantiate this advantage. As it can be found, the raw cell embeddings often tend to mix the drug sensitivity property to a specific drug. However, the cell embeddings generated by our \method{} illustrate the discernibility of cell sensitivity to various drugs. Cells with similar sensitivities typically aggregate together, whereas cells with distinct sensitivities are dispersed throughout.
The confusion matrix results in Fig.~\ref{fig: dsp} (d) reveal near-perfect sensitivity prediction accuracy, with only minor degradation in resistance classification. Overall, \method{} significantly surpasses other specialist and generalist models in drug sensitivity prediction.

\section{Conclusion and Limitations}

Here we identify the shortcomings of both pretrain-based and LLM-based scFMs, and propose a double-win framework termed \method{} that excels in both cost-effectiveness and high-performance for versatile single-cell analysis. The core of \method{} is generating bi-level biological knowledge, which is retrieved as additional supervision signals during the pretraining phase. Remarkably, \method{} outperforms cutting-edge scFMs across multiple downstream tasks. 

We also admit limitations. The power of \method{} is depended by the choices of LLMs to produce embeddings and descriptions, which mean that we need to track the development very frequently. Powerful LLMs are also closed-source, and thus using open-source LLMs for further exploration should be encouraged. In future work, we aim to further explore scaling laws for single-cell foundation models, with the goal of developing a unified, all-in-one framework.

\section{Credit Taxonomy}

F.Z. and T.L. proposed this study. T.L. and F.Z. designed the method. T.L., F.Z., J.C. performed experiments. All authors wrote and reviewed the manuscript. T.L. and Y.Z. supervised this study.

\bibliographystyle{natbib}
\bibliography{scibib}

\appendix

\section{Introduction to Incorporated Datasets}
\paragraph{CITE-seq and ASAP-seq PBMC dataset.}
The CITE-seq and ASAP-seq PBMC datasets contain multimodal data from both control and stimulated conditions. After preprocessing, 4,644 CITE-seq cells and 4,502 ASAP-seq cells associated with 17,441 genes are included in the experiments. The CITE-seq data includes seven categories: B cells, Effector CD4+ T cells, Effector CD8+ T cells, Monocytes cells, NK cells, Naive CD4+ T cells, and Naive CD8+ T cells. The ASAP-seq data includes nine categories: B cells, DC cells, Effector CD4+ T cells, Effector CD8+ T cells, Monocytes cells, NK cells, Naive CD4+ T cells, Naive CD8+ T cells, and others. This dataset can be downloaded from GSE156478 (\url{https://www.ncbi.nlm.nih.gov/geo/query/acc.cgi?acc=GSE156478}). 

\paragraph{Liver dataset.} We create a Liver dataset with two batches from two different studies. We manually unify the cell type labels of two datasets by regular expressions to remove the mismatched spelling. We use the same Scanpy pre-processing step to normalize gene expression, unify gene names and create two batches including MacParland and Aizarani for training/testing. We have considered both directions.


\paragraph{PBMC 10K dataset.}
The PBMC 10K dataset includes two batches of scRNA-seq data from human PBMCs of a healthy donor. After preprocessing, this dataset features 3,346 differentially expressed genes. The first batch and the second batch comprise 7,982 cells and 4,008 cells, respectively. Cell groups were labeled by Seurat~\cite{satija2015spatial} and categorized into nine types: B cells, CD4+ T cells, CD8+ T cells, CD14+ monocytes, dendritic cells, NK cells, FCGR3A+ monocytes, megakaryocytes, and others. This dataset can be downloaded from the scVI tools~\cite{virshup2023scverse,gayoso2022python} (\url{https://scvi-tools.org/}) using the API scvi.data.pbmc\_dataset.

\paragraph{BMMC dataset.}
The BMMC dataset uses the CITE-seq protocol and comprises paired measurements of scRNA-seq and protein abundance in BMMCs. This dataset includes cells from 12 healthy human donors, organized into 12 batches. After preprocessing, the data encompass 90,261 cells and each cell contains 13,953 genes and 134 surface proteins. There are 45 distinct cell types in this dataset. This dataset can be downloaded from GSE194122 (\url{https://www.ncbi.nlm.nih.gov/geo/query/acc.cgi?acc=GSE194122}).

\paragraph{GSE149383-Erlotinib.}
The GSE149383 dataset comprises 2,739 human lung cancer cells and 8,380 associated genes, along with their respective sensitivity properties to the drug Erlotinib. This dataset can be downloaded from GSE149383 (\url{https://www.ncbi.nlm.nih.gov/geo/query/acc.cgi?acc=GSE149383}).

\paragraph{GSE117872-Cisplatin.}
The GSE117872 dataset comprises 1,302 human oral squamous cancer cells and 22,744 associated genes, along with their respective sensitivity properties to the drug Cisplatin. This dataset can be downloaded from GSE117872 (\url{https://www.ncbi.nlm.nih.gov/geo/query/acc.cgi?acc=GSE117872}).

\section{Introduction to Compared Methods}

To evaluate the performance of \method{} for downstream tasks, we benchmarked our method against a range of state-of-the-art (SOTA) approaches. For tasks like cell type annotation, batch effect correction, and multi-omics data integration, we compared \method{} with four scFMs, including scBERT, scGPT, GenePT, and scELMo. For the rare cell type identification task, we employed scCAD as the framework and replaced the original cell embeddings with cell representations obtained by our approach. For the drug sensitivity prediction task, we compared \method{} with both specialist models like vanilla Transformer, scDEAL, and generalist models like GPT-3.5.

\paragraph{scBERT \cite{yang2022scbert}.} 
scBERT is a pretrain-based scFM that utilizes a Performer cell encoder and employs masked gene modeling for pretraining on one million cells from the Panglao dataset~\cite{franzen2019panglaodb}. We used the pretrained model provided by the authors from (\url{https://drive.weixin.qq.com/s?k=AJEAIQdfAAoUxhXE7r#}) and validated downstream tasks according to their publicly available code from (\url{https://github.com/TencentAILabHealthcare/scBERT}). All other settings were kept consistent with those provided by the authors. 

\paragraph{scGPT \cite{cui2024scgpt}.}
scGPT is another pretrain-based scFM that employs a decoder-based architecture combined with attention mask for autoregressive pretraining on 33 million cells. The authors provided several pretrained model weights, with pretraining data ranging from 1.8 million heart cells to 33 million normal human cells. We utilized the model pretrained on the largest dataset from (\url{https://drive.google.com/drive/folders/1oWh_-ZRdhtoGQ2Fw24HP41FgLoomVo-y}) and conducted downstream task experiments following the authors' publicly available code from (\url{https://github.com/bowang-lab/scGPT}). All other settings were kept consistent with those provided by the authors.

\paragraph{Geneformer \cite{theodoris2023transfer}.} Geneformer is a pretrain-based scFM that uses a transformer encoder architecture for context-aware gene representation learning from rank-value encoded single-cell transcriptomes. The model was pretrained on approximately 30 million human single-cell transcriptomes from the Genecorpus-30M dataset, enabling transfer to diverse downstream tasks such as cell type annotation, perturbation prediction, and gene network analysis. The authors provided pretrained model weights through Hugging Face, and we utilized the publicly available pretrained Geneformer model from (\url{https://huggingface.co/ctheodoris/Geneformer}) for downstream task experiments.

\paragraph{scFoundation \cite{hao2024large}.} scFoundation is a pretrain-based scFM that employs a transformer-based architecture to learn universal cell representations from large-scale single-cell gene expression profiles. The model was pretrained on over 50 million human single-cell transcriptomes, enabling broad transferability across downstream tasks such as cell type annotation, perturbation response prediction, and gene expression reconstruction. The authors provided pretrained model weights, and we utilized the publicly available pretrained scFoundation model from (\url{https://github.com/biomap-research/scFoundation}) for downstream task experiments.

\paragraph{UCE \cite{rosen2023universal}.} UCE is a pretrain-based scFM that employs a large transformer-based architecture to generate universal cell embeddings for single-cell transcriptomic data. The model was pretrained in a self-supervised manner on more than 36 million cells from over 300 datasets across human and seven other species, aiming to learn a unified biological latent space across tissues, datasets, and species. A key feature of UCE is that it tokenizes genes using protein-language-model-derived embeddings, enabling cross-species and even new-species generalization. The authors provided pretrained model weights and released the implementation publicly, and we utilized the pretrained UCE model from (\url{https://github.com/snap-stanford/UCE}) for downstream task experiments.

\paragraph{Cell2Sentence \cite{levine2023cell2sentence}.} Cell2Sentence is a pretrain-based scFM that reframes single-cell transcriptomes as natural-language-like sentences where genes are ordered according to their expression ranks and processed using large language models. Instead of directly modeling continuous gene expression matrices, Cell2Sentence converts each cell into a sequence of gene tokens, enabling decoder-based language models to learn gene-gene context and generate biologically meaningful cell representations. The authors provided pretrained C2S models and tutorials for rank transformation, reconstruction, and downstream analysis, and we utilized the publicly available implementation from (\url{https://github.com/vandijklab/cell2sentence}) for downstream task experiments.

\paragraph{GenePT \cite{chen2023genept}.}
GenePT is an LLM-based scFM that obtains text embeddings of features (such as genes and proteins) from LLM using text descriptions from the NCBI database as prompts. It then directly uses these LLM embeddings to construct cell embeddings. We employed the GenePT-w method, as provided by the authors from (\url{https://github.com/yiqunchen/GenePT}), to construct cell embeddings and validate them for downstream tasks. All other settings were kept consistent with those provided by the authors.

\paragraph{scELMo \cite{liu2023scelmo}.}
scELMo is another LLM-based scFM, which can be considered as an improved version of GenePT. It obtains embeddings through enhanced prompts to construct high-quality cell embeddings. We utilized the code provided by the authors from (\url{https://github.com/HelloWorldLTY/scELMo}) to build cell embeddings and validate them for downstream tasks. All other settings were kept consistent with those provided by the authors.

\paragraph{scCAD \cite{xu2024sccad}.}
scCAD is a SOTA algorithm for rare cell type identification. It employs the PCA method for clustering to obtain cell embeddings. We conducted experiments using the publicly available code from (\url{https://github.com/xuyp-csu/scCAD}) and replaced the original cell embeddings with representations from our framework. Experimental analysis was performed using the cell embeddings generated by our approach. All other parameters of the algorithm were kept consistent with the original implementation.

\paragraph{Transformer \cite{vaswani2017attention}.}
Transformer is a famous architecture and was initially designed for natural language processing tasks. It introduces self-attention mechanisms and provides standards for many other fields, including vision, audio, and scientific research. For the drug sensitivity prediction task, all settings and parameters were kept consistent with the standard implementation.

\paragraph{scDEAL \cite{chen2022deep}.}
scDEAL is a specialist model designed for the drug sensitivity prediction task. It employs a neural network and tries to establish connections between genes and drug responses.
The source code and implementation can be found at (\url{https://github.com/OSU-BMBL/scDEAL}). All settings and parameters were kept consistent with the original implementation.

\paragraph{LangCell \cite{langcell}.}
LangCell is a cell-text pretrained model that couples a Geneformer-style cell encoder with a text encoder and aligns the two through contrastive learning over cell-type descriptions, so that a cell can be annotated by matching it to type descriptions rather than to a fixed label set. Of the published methods it is the closest in design to RAGCell, since both introduce textual descriptions of cell identity into representation learning, and it is therefore the most informative comparison for isolating what the retrieval component contributes (\url{https://github.com/PharMolix/LangCell}).

\paragraph{GPT-3.5 and GPT-4o mini.}
GPT-3.5 is a reliable LLM developed by OpenAI. Based on the GPT-3 architecture, GPT-3.5 can understand and generate human-liked text, which could be employed for many tasks and applications, including the drug sensitivity prediction task. The use of GPT-3.5 API can be found at (\url{https://platform.openai.com/docs/models/gpt-3.5-turbo}). We also test the usage of GPT-4o mini in our ablation studies. API: \url{https://platform.openai.com/docs/models/gpt-4o-mini}

\section{Implementation Details}
\method{} employs the patch-based Transformer model as the cell model. Each cell is split into 128 patches for data tokenization.
For cell-level knowledge databases, we employ the GPT-4o mini model to generate function descriptions and use the `text-embedding-3-large' model to extract embeddings, with an embedding dimension of 3072.
For feature-level knowledge databases, we leverage the GPT-3.5 model to generate function descriptions and use the `text-embedding-ada-002' model to extract embeddings, with an embedding dimension of 1536.
To map cell and text representations into a shared space, we use two MLPs as projection layers, reducing both representations to a low-dimensional space of 128 dimensions.
All experiments are conducted using Pytorch with 80GB A100 GPUs as support. For cell-text alignment, we set the batch size to 256 and trained the cell model with the AdamW optimizer for 100 epochs. 
For finetuning tasks, we randomly split the datasets into 80\% for training, 10\% for validation, and 10\% for testing.
For zero-shot tasks, we ensure that there is no data leakage or overlap during the pretraining phase.
The learning rate is set to 1.5e-4 during the pretraining stage and 1e-4 during the finetuning stage, respectively.

\section{Evaluation Metrics}

\paragraph{Cell Type Annotation.}
For the cell type annotation task, we evaluated the model's performance using four common metrics for classification tasks: Accuracy, Precision, Recall, and F1 score. The calculation processes for each metric are as follows:
\begin{equation}
     \text{Accuracy} = \frac{\text{TP} + \text{TN}}{\text{TP} + \text{TN} + \text{FP} + \text{FN}},
\end{equation}
\begin{equation}
    \text{Precision} = \frac{1}{N} \sum_{i=1}^{N}  \frac{\text{TP}_i}{\text{TP}_i + \text{FP}_i},
\end{equation}
\begin{equation}
    \text{Recall} = \frac{1}{N} \sum_{i=1}^{N} \frac{\text{TP}_i}{\text{TP}_i + \text{FN}_i},
\end{equation}
\begin{equation}
    \text{F1 Score} = \frac{1}{N} \sum_{i=1}^{N} \frac{2 \times \text{Precision}_i \times \text{Recall}_i}{\text{Precision}_i + \text{Recall}_i},
\end{equation}
where $\text{TP}$, $\text{TN}$, $\text{FP}$, and $\text{FN}$ are short for true positives, true negatives, false positives, and false negatives, respectively. $N$ denotes the number of samples per cell type.

\paragraph{Batch Effect Correction.}
For the batch effect correction task, we evaluated the model's performance using several common cell clustering metrics, specifically normalized mutual information (NMI), adjusted rand index (ARI), and average silhouette width (ASW). NMI can be calculated as:
\begin{equation}
    \text{I}(U, V) = \sum_{i=1}^{|U|} \sum_{j=1}^{|V|} P(i, j) \log \left( \frac{P(i, j)}{P(i) P(j)} \right),
\end{equation}
\begin{equation}
    \text{NMI}(U, V) = \frac{I(U, V)}{\text{mean}(\text{H}(U), \text{H}(V))},
\end{equation}
where $P (i,j)$ is the probability that the $i$-th cluster in clustering $U$ and the $j$-th cluster in clustering $V$ occur simultaneously.
$P(i)$ and $P(j)$ are the probabilities of the $i$-th cluster in $U$ and the $j$-th cluster in $V$, respectively.

ARI can be calculated as:
\begin{equation}
    \text{RI} = \frac{\text{TP} + \text{TN}}{\text{TP} + \text{TN} + \text{FP} + \text{FN}},
\end{equation}
\begin{equation}
    \text{ARI} = \frac{\text{RI} - E[\text{RI}]}{\max(\text{RI}) - E[\text{RI}]},
\end{equation}
where $\text{TP}$, $\text{TN}$, $\text{FP}$, and $\text{FN}$ are short for true positives, true negatives, false positives, and false negatives, respectively. $E[RI]$ is the expected RI of random labeling.

The calculation process for ASW can be represented as follows:
\begin{equation}
    a(i) = \frac{1}{|C_i| - 1} \sum_{j \in C_i, j \neq i} d(i, j),
\end{equation}
\begin{equation}
    b(i) = \min_{k \neq i} \frac{1}{|C_k|} \sum_{j \in C_k} d(i, j),
\end{equation}
\begin{equation}
    s(i) = \frac{b(i) - a(i)}{\max\{a(i), b(i)\}},
\end{equation}
\begin{equation}
    \text{ASW} = \frac{1}{N} \sum_{i=1}^{N} s(i),
\end{equation}
where $C_i$ and $C_k$ are clusters containing sample points $i$ and $k$, and $d(i, j)$ is the distance between sample points $i$ and $j$. $N$ denotes the number of cell samples.

\paragraph{Multi-omics Data Integration.}
For the multi-omics data integration task, we employed three metrics, namely NMI, ARI, and ASW, to evaluate the performance of the models. The calculation processes for these metrics have been previously described. 

\paragraph{Rare Cell Type Identification.}
For the rare cell type identification task, we employed three metrics, namely Precision, Recall, and F1 score, to evaluate the performance of the models. The calculation processes for these metrics have been previously described.

\paragraph{Drug Sensitivity Prediction.}
For the drug sensitivity prediction task, we employed four metrics, namely Precision, Recall, F1 score, and Accuracy, to evaluate the performance of the models. The calculation processes for these metrics have been previously described.

\begin{figure*}[t]
    \centering
    \includegraphics[width=\linewidth]{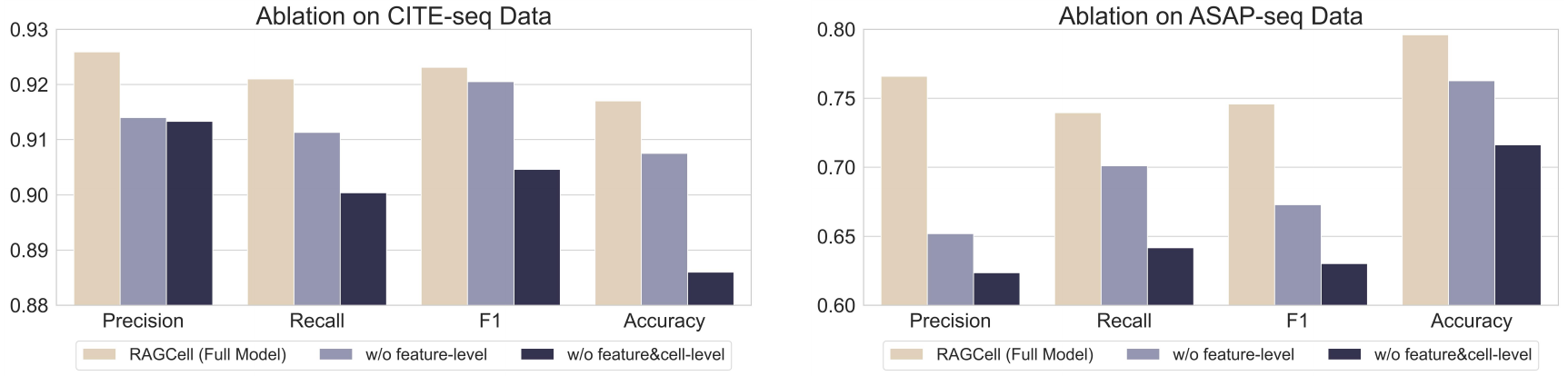}
    \caption{Ablation study on CITE-seq and ASAP-seq data.}
    \label{fig: ablation}
\end{figure*}
\begin{figure*}[th]
    \centering
    \includegraphics[width=\linewidth]{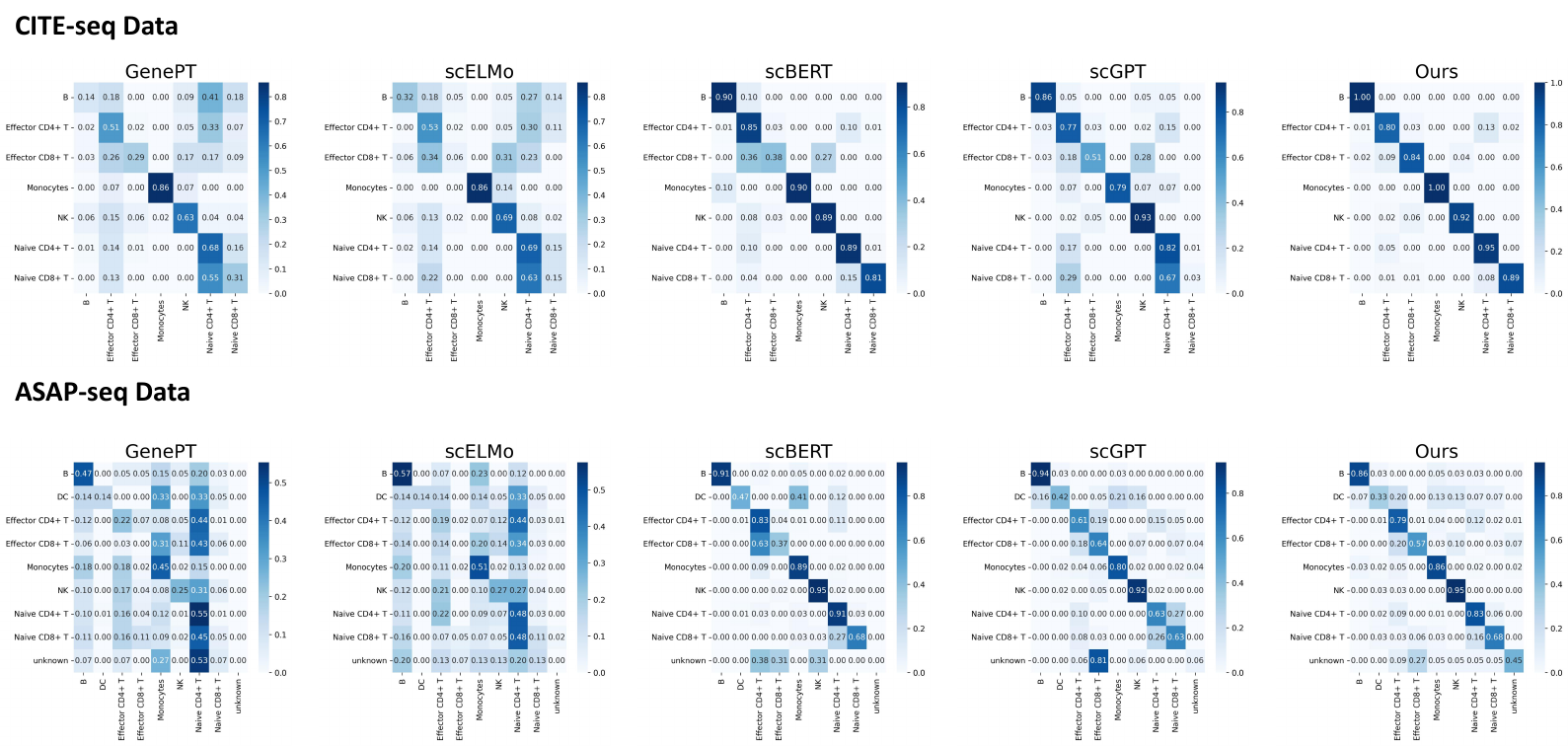}
    \caption{Confusion matrix results on CITE-seq and ASAP-seq data.}
    \label{fig: cm}
\end{figure*}

\section{Additional Experiments}


\subsection{Ablation Study}

To verify the effectiveness of incorporating priors from LLMs, we also provide some model variants by gradually removing text descriptions at each level, with results shown in Fig.~\ref{fig: ablation}.
As illustrated in the figure, `w/o feature-level' signifies that only cell-level descriptions were included, whereas `w/o feature\&cell-level' indicates the absence of all textual information.
The results indicate performance declines among various model variants, which suggests the need of both cell-level and feature-level text descriptions in enhancing cell representations.

\subsection{Additional Cell Type Annotation Results}
We also present the cell type annotation confusion matrix results in Fig.~\ref{fig: cm}, which highlights the superior performance of \method{} compared to other scFMs. \method{} achieves significantly higher annotation accuracy consistently across diverse cell types.

\begin{figure*}[h]
    \centering
    \includegraphics[width=0.8\linewidth]{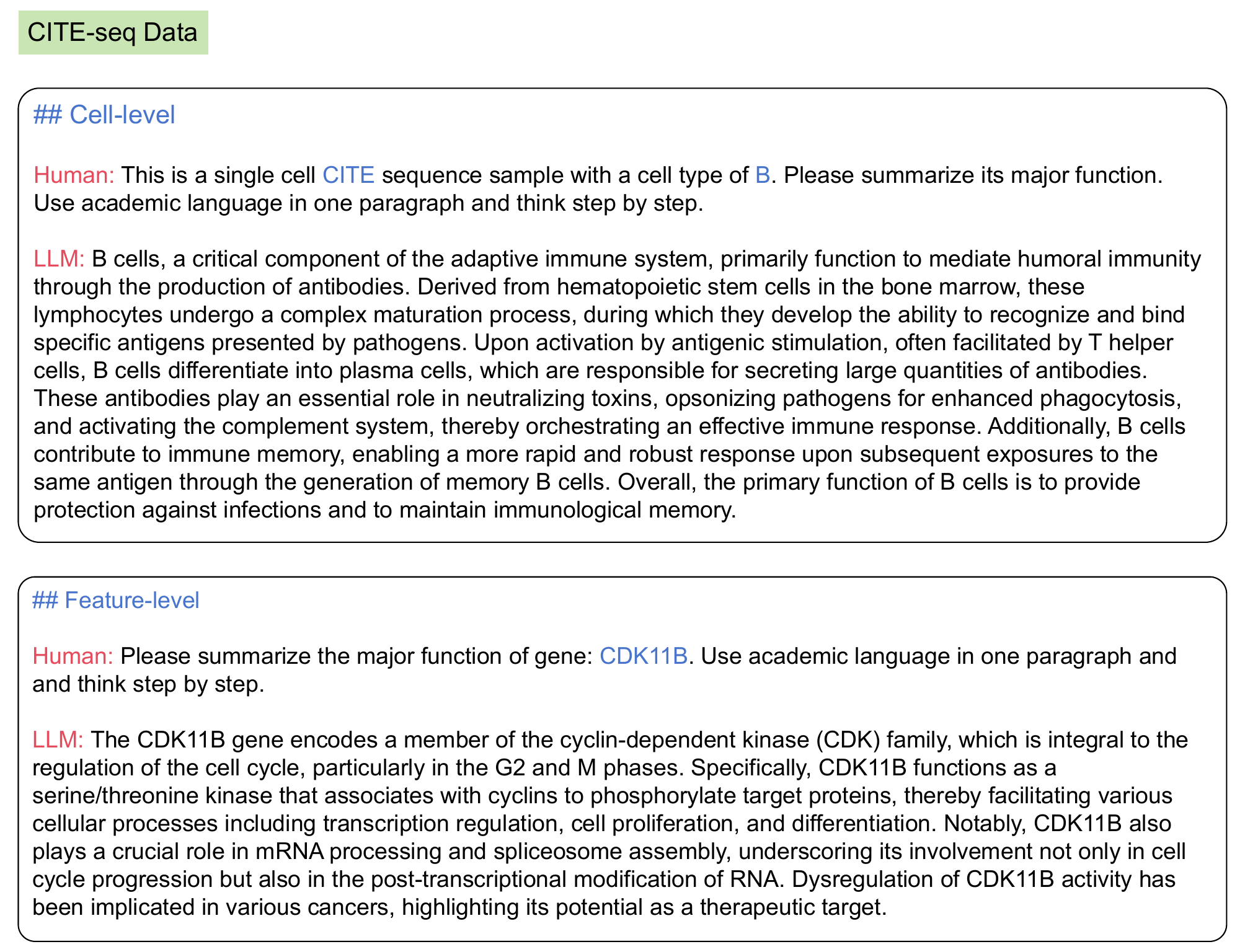}
    \caption{Examples of cell-level and feature-level (using gene as an example) text descriptions on CITE-seq data.}
    \label{fig: case_cite}
\end{figure*}

\begin{figure*}[h]
    \centering
    \includegraphics[width=0.8\linewidth]{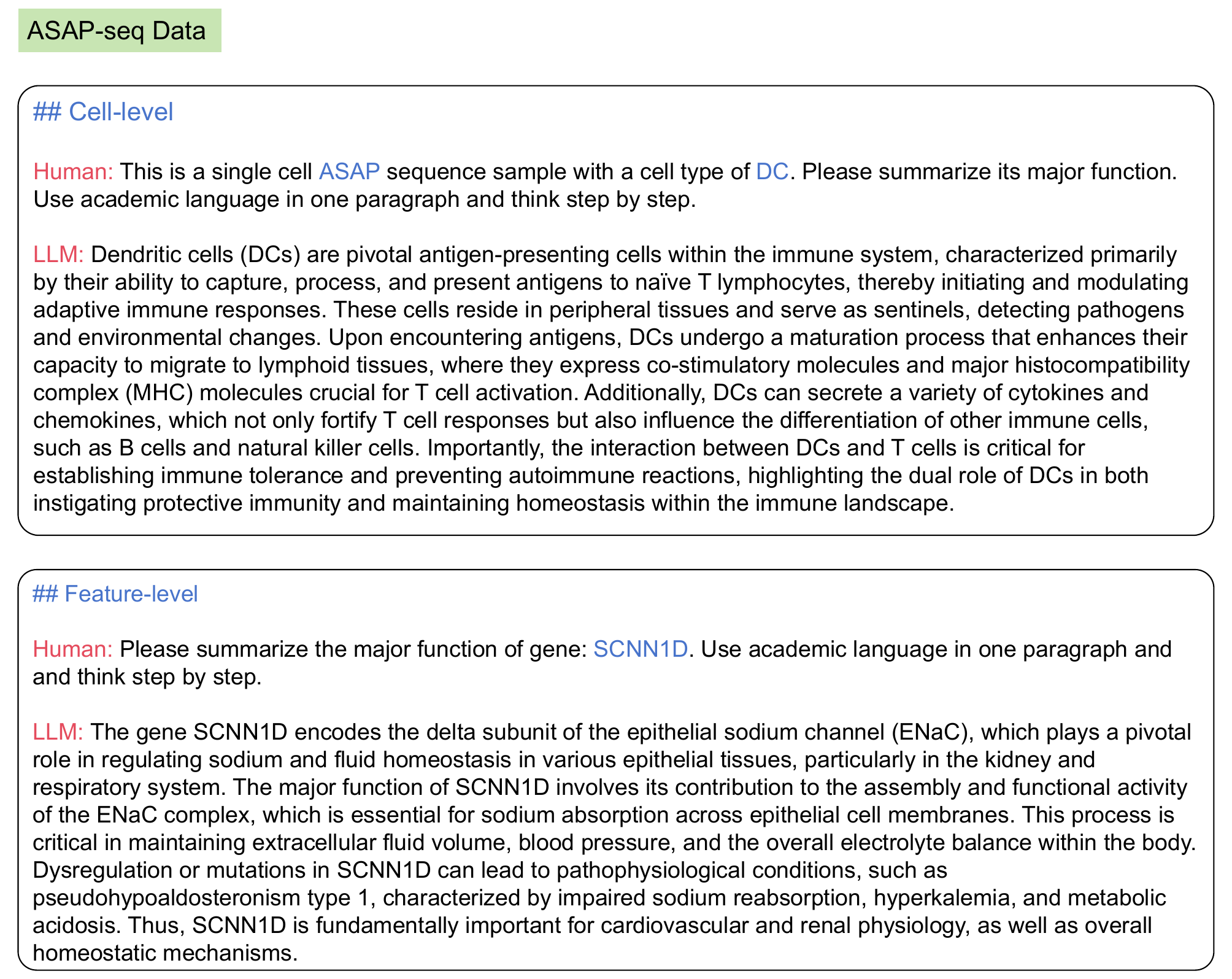}
    \caption{Examples of cell-level and feature-level (using gene as an example) text descriptions on ASAP-seq data.}
    \label{fig: case_asap}
\end{figure*}

\begin{figure*}[h]
    \centering
    \includegraphics[width=0.8\linewidth]{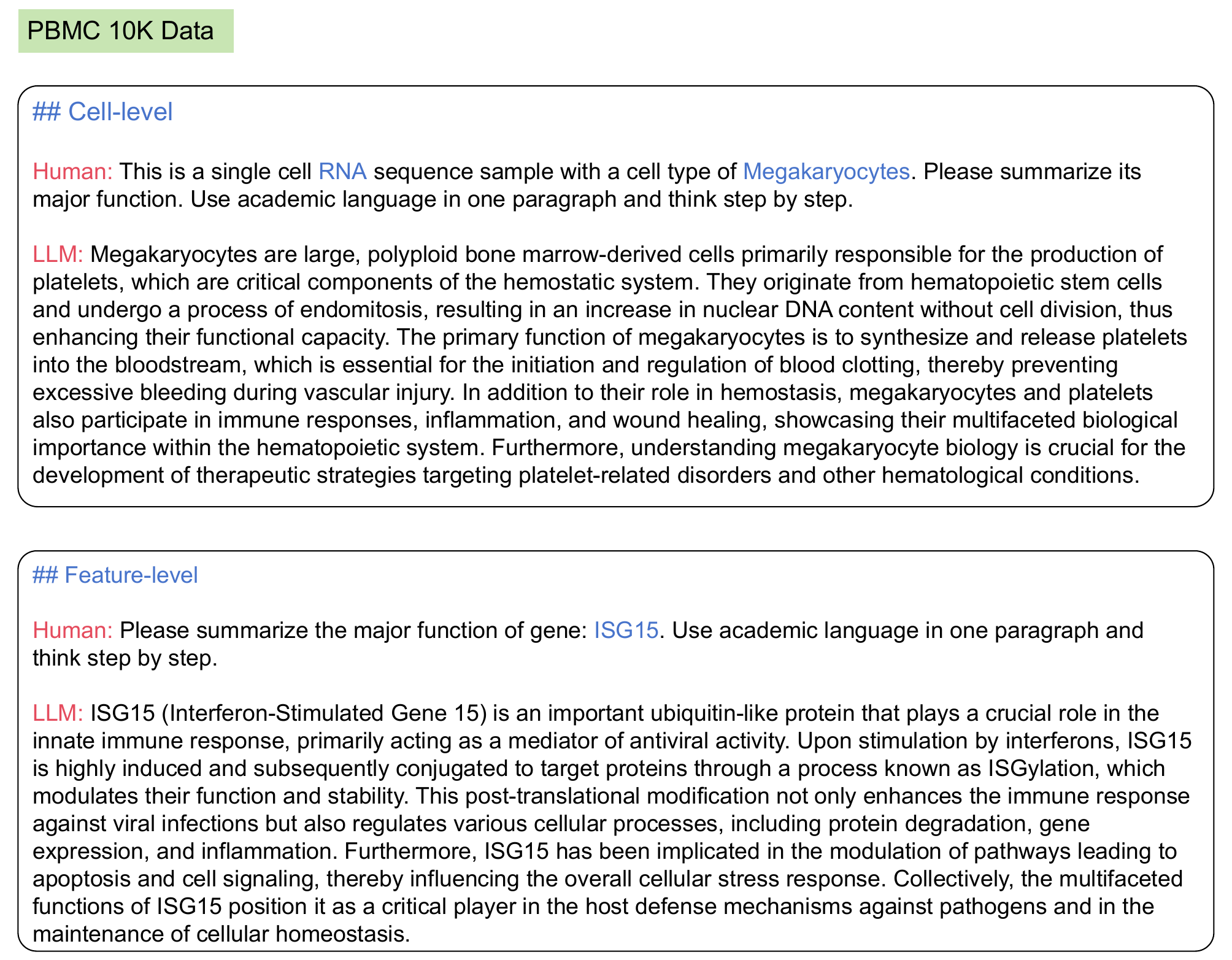}
    \caption{Examples of cell-level and feature-level text descriptions on PBMC 10K data.}
    \label{fig: case_pbmc}
\end{figure*}

\begin{figure*}[h]
    \centering
    \includegraphics[width=0.8\linewidth]{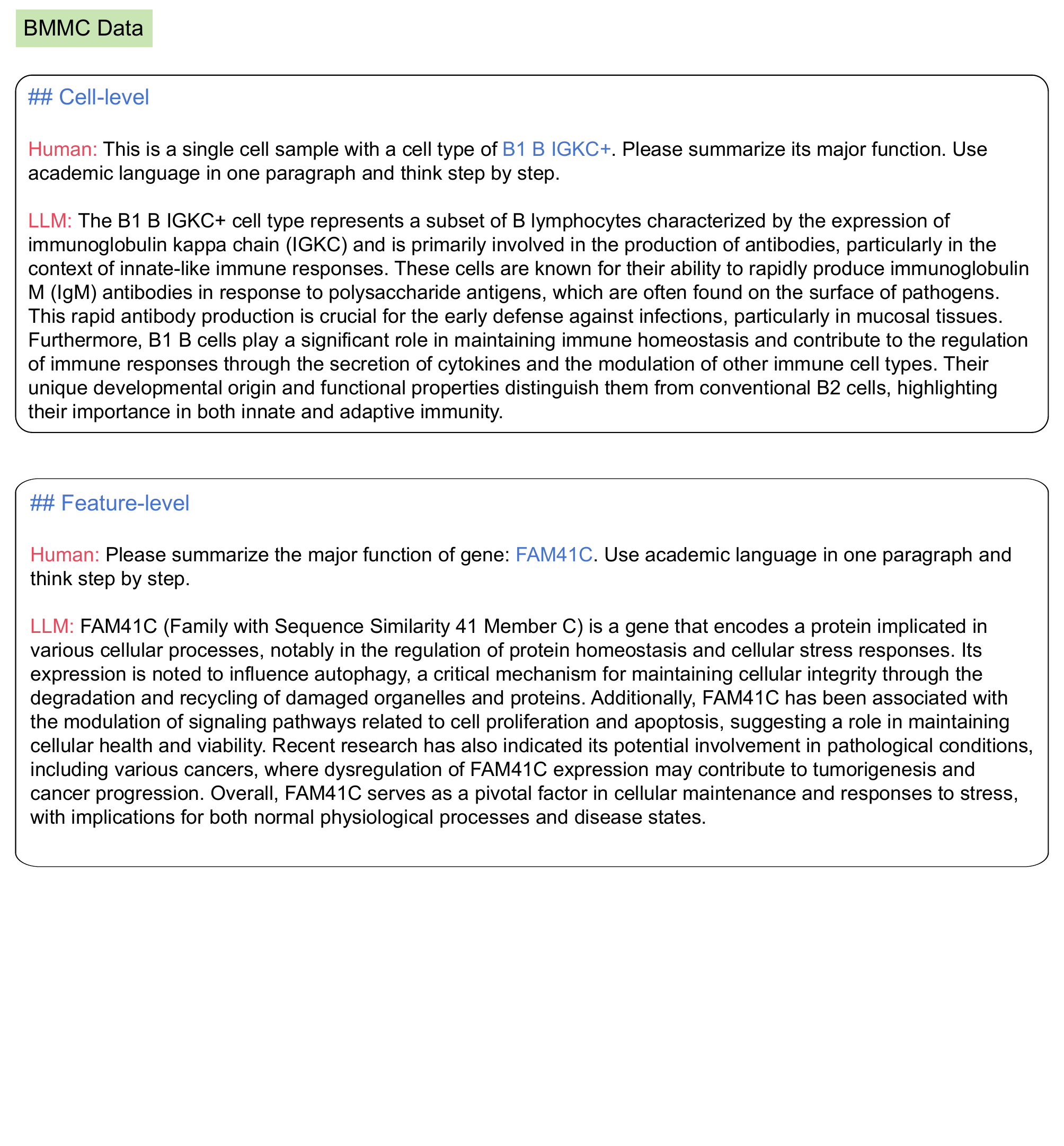}
    \caption{Examples of cell-level and feature-level text descriptions on BMMC data.}
    \label{fig: case_bmmc}
\end{figure*}

\begin{figure*}[h]
    \centering
    \includegraphics[width=0.8\linewidth]{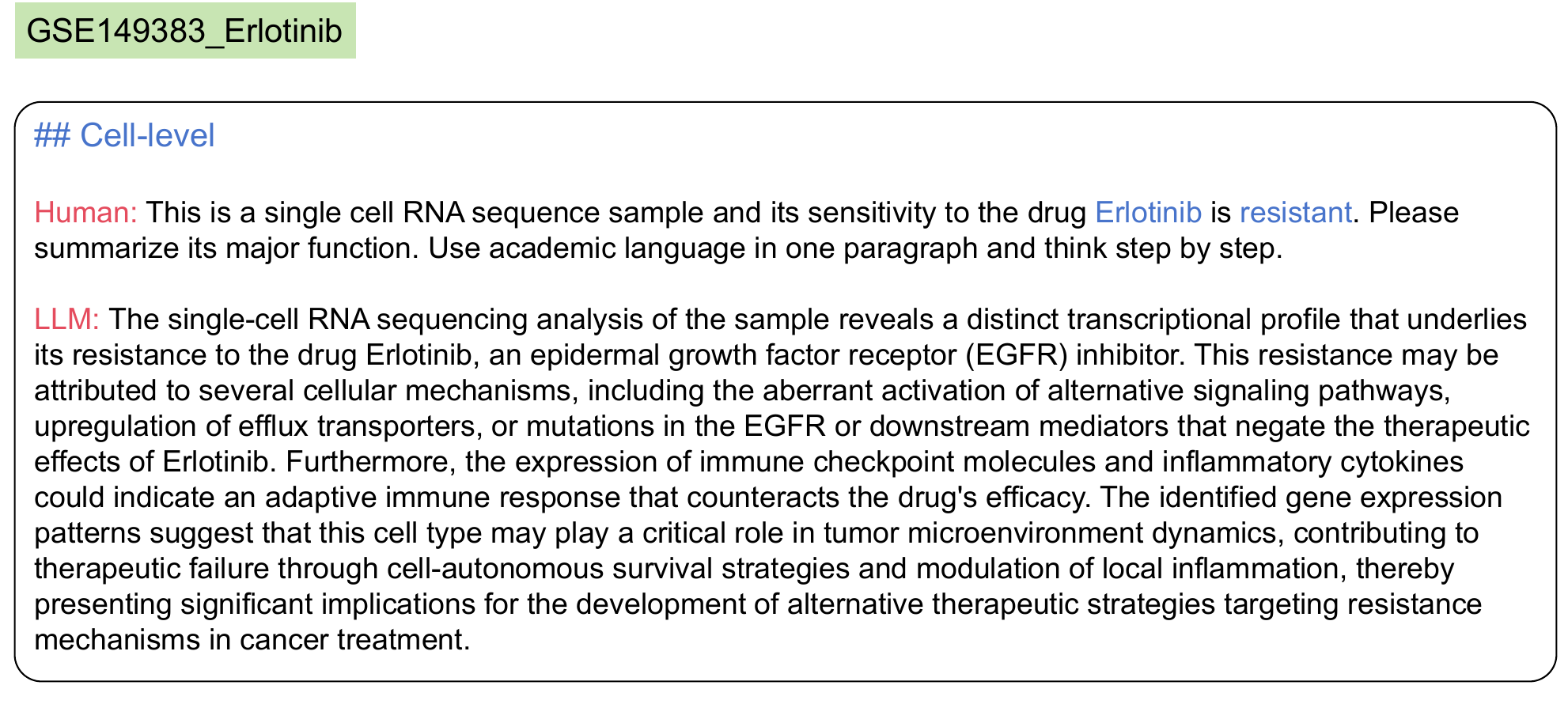}
    \caption{Examples of cell-level and feature-level text descriptions on GSE149383 data.}
    \label{fig: case_GSE149383}
\end{figure*}

\begin{figure*}[h]
    \centering
    \includegraphics[width=0.8\linewidth]{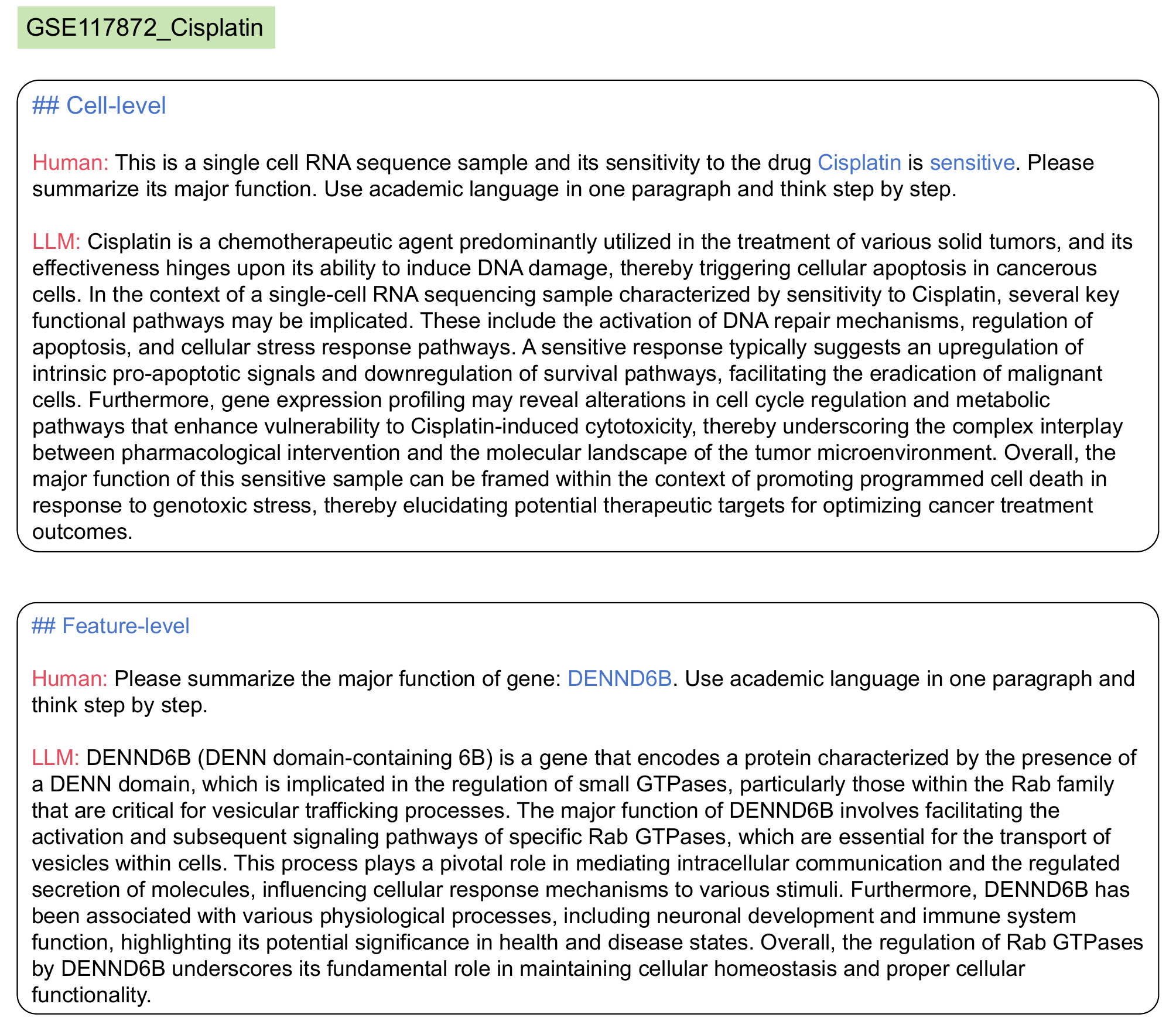}
    \caption{Examples of cell-level and feature-level text descriptions on GSE117872 data.}
    \label{fig: case_GSE117872}
\end{figure*}

\section{Examples of Text Descriptions}

Here we provide some examples of cell-level and feature-level text descriptions on different datasets in Fig.~\ref{fig: case_cite}, Fig.~\ref{fig: case_asap}, Fig.~\ref{fig: case_pbmc}, Fig.~\ref{fig: case_bmmc}, Fig.~\ref{fig: case_GSE149383}, and Fig.~\ref{fig: case_GSE117872}.

\subsection{Additional experiments.}

\begin{figure*}[h]
    \centering
    \includegraphics[width=0.8\linewidth]{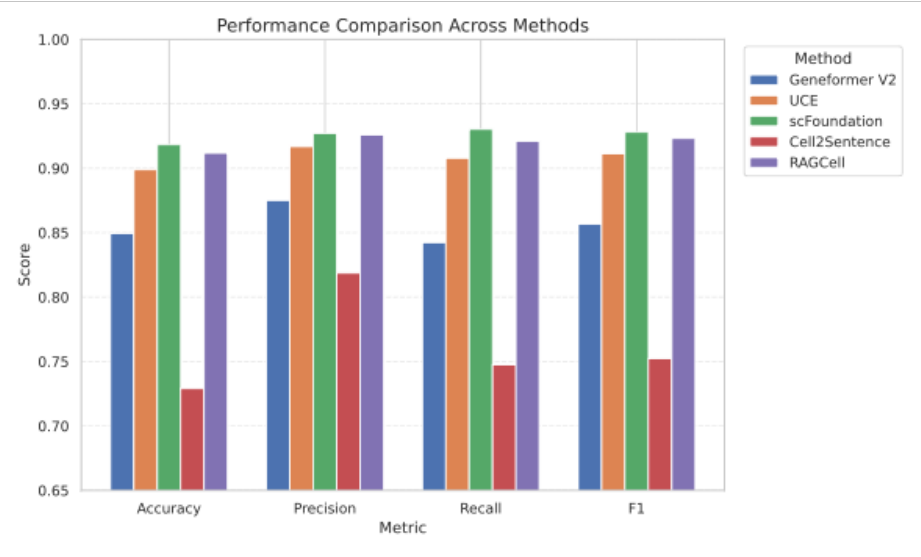}
    \caption{Benchmarking results with more methods for cell-type annotation.}
    \label{fig: morebench_asap}
\end{figure*}

\begin{figure*}[h]
    \centering
    \includegraphics[width=0.8\linewidth]{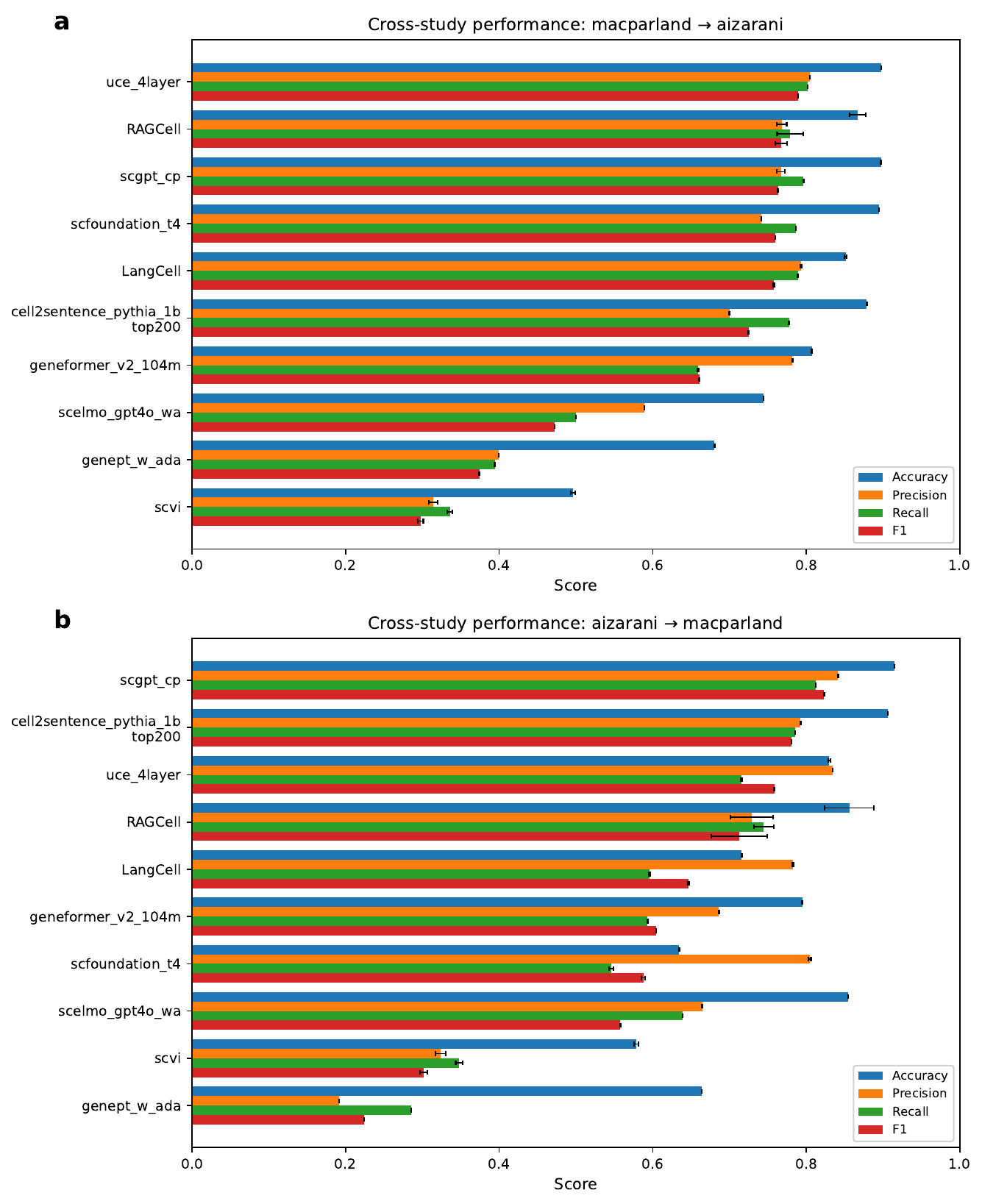}
    \caption{Benchmarking results with more methods for cell-type annotation based on the Liver dataset (cross-study comparison and 5 seeds update).}
    \label{fig: morebench_corssdata}
\end{figure*}

\begin{figure*}[h]
    \centering
    \includegraphics[width=0.8\linewidth]{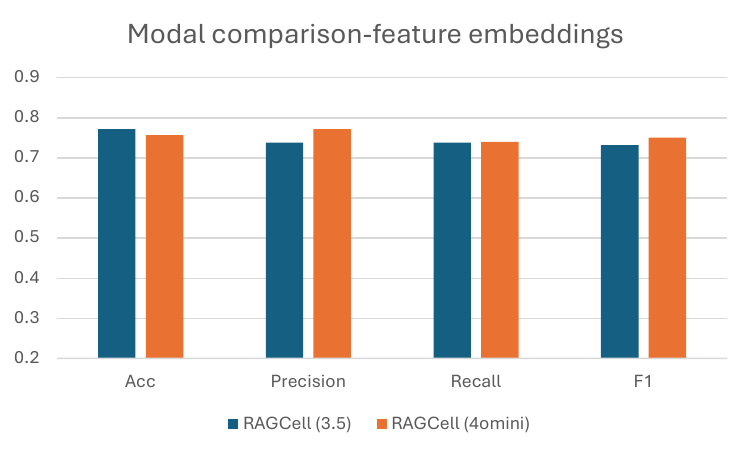}
    \caption{Ablation studies for embedding model selection.}
    \label{fig: morebench_gpt}
\end{figure*}

\begin{figure*}[h]
    \centering
    \includegraphics[width=0.8\linewidth]{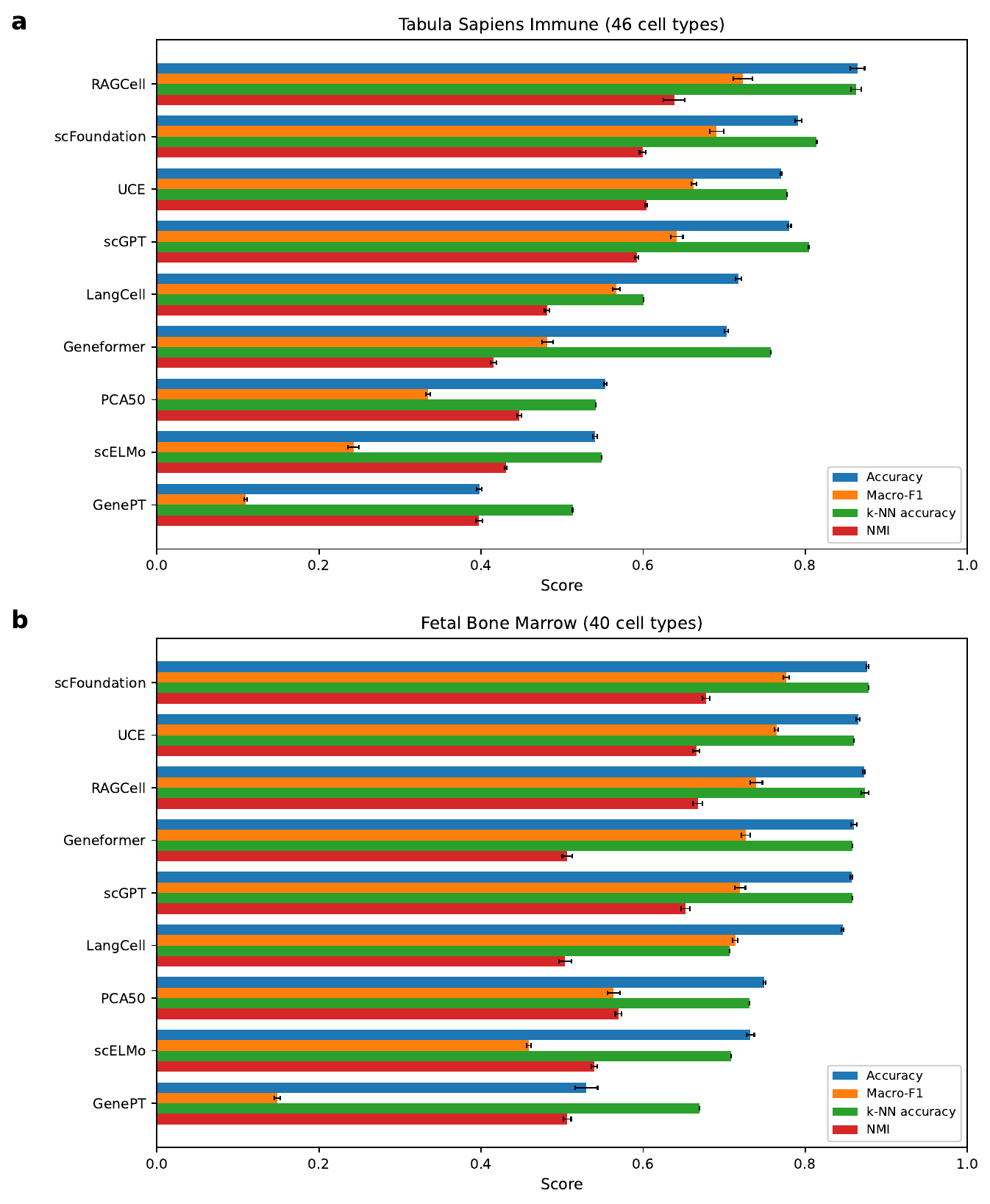}
    \caption{CTA results on two additional atlases. a, Tabula Sapiens Immune. b, Fetal Bone Marrow. Bars are means over five seeds; error bars are standard deviations.}
    \label{fig: largescale}
\end{figure*}

\begin{figure*}[h]
    \centering
    \includegraphics[width=0.8\linewidth]{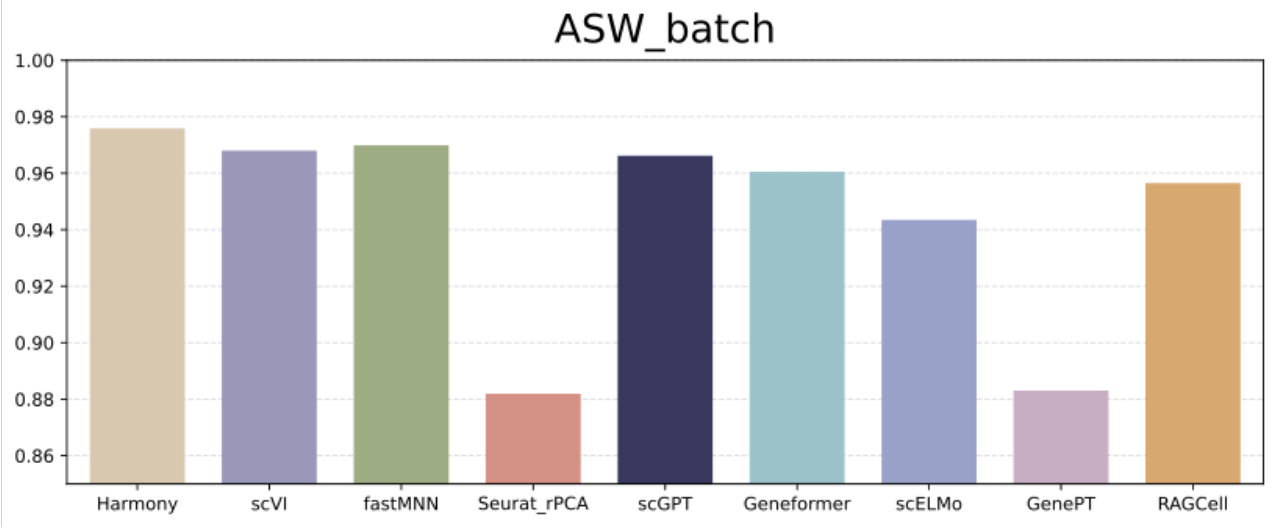}
    \caption{ASW score comparison for batch effect correction with more baselines.}
    \label{fig: morebench_bec}
\end{figure*}

\begin{figure*}[h]
    \centering
    \includegraphics[width=0.8\linewidth]{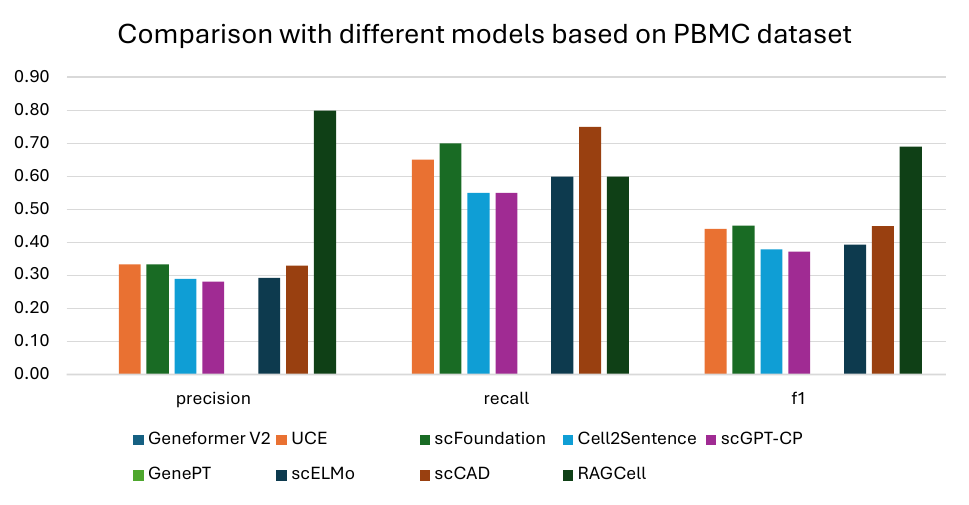}
    \caption{A more comprehensive comparison for rare cell type identification.}
    \label{fig: morebench_rarecelltypedetect}
\end{figure*}

\end{document}